\documentclass[aps,pra,10pt,twocolumn,numerical]{revtex4-1}
\usepackage[utf8x]{inputenc}
\usepackage{amssymb,amsmath}
\usepackage{multirow}
\usepackage{epsfig}
\usepackage{hyperref}
\usepackage{subfigure}
\usepackage{bbm}
\usepackage{stmaryrd}
\usepackage{pspicture}
\usepackage{natbib}
\usepackage[english]{babel}
\usepackage{epstopdf}
\usepackage{graphics}
\usepackage{graphicx}

\usepackage{hyperref}
\hypersetup{colorlinks=true, breaklinks=true, linkcolor=blue, citecolor=red}

\newcommand{\ket}[1]{\left|#1\right>}
\newcommand{\bra}[1]{\left<#1\right|}
\newcommand{\braket}[1]{\left<#1\right>}

\newcommand{\op}[1]{\hat{#1}^{\phantom{\dagger}}}
\newcommand{\opd}[1]{\hat{#1}^{\dagger}}
\newcommand{\tr}[1]{{\rm{Tr}}\left\{#1\right\}}
\newcommand{\ie}{i\@.e\@.}
\newcommand{\eg}{e\@.g\@.}

\newcommand{\ii}{{\rm{i}}\,}

\newcommand{\equ}[1]{\begin{align}{#1}\end{align}}

\newcommand{\rklamm}[1]{ \left( #1 \right) }
\newcommand{\eklamm}[1]{ \left[ #1 \right] }
\newcommand{\sklamm}[1]{ \left\{ #1 \right\} }

\newcommand{\bint}[3]{\int\limits_{#1}^{#2} d{#3} \, }

\newcommand{\figref}[1]{Fig.~\ref{#1}}
\newcommand{\secref}[1]{Sec.~\ref{#1}}
\newcommand{\equref}[1]{Eq.~\eqref{#1}}

\begin{document}
\title{Transport with ultracold atoms at constant density}
\author{Christian Nietner} \email{cnietner@itp.tu-berlin.de}
\author{Gernot Schaller}
\author{Tobias Brandes}
\affiliation{Institut f\"ur Theoretische Physik, Technische Universit\"at Berlin, Hardenbergstr. 36, 10623 Berlin, Germany}
%
%
%
%
\begin{abstract}
   We investigate the transport through a few-level quantum system described by a Markovian master equation with temperature and particle-density-dependent chemical potentials. From the corresponding Onsager relations we extract linear response transport coefficients in analogy to the electronic conductance, thermal conductance, and thermopower. Considering ideal Fermi and Bose gas reservoirs, we observe steady-state currents against the thermal bias as a result of the nonlinearities introduced by the constraint of a constant particle density in the reservoirs. Most importantly, we find signatures of the on-set of Bose-Einstein condensation in the transport coefficients.
\end{abstract}
\maketitle
%
%
\section{Introduction}
%
%
Transport processes are widespread and commonly occur in many fields of physics, chemistry, and biology. In electronic and photonic systems, the chemical potential as well as the temperature are normally treated as independently controllable parameters \cite{Kouwenhoven2001,Brandes2005,Nietner2012b}. 

In equilibrium thermodynamics, the chemical potential is in general a function of the intensive properties of the gas such as, \eg, temperature and particle density. In most setups studied so far, assuming an independent, constant chemical potential is a valid conjecture, since the transport setup is usually embedded in a much larger environment. This surrounding environment acts as a particle and temperature reservoir providing the necessary resources to fix the chemical potential at a constant value. 

However, there are also transport setups which can not be treated in this manner. In particular, there has been a lot of progress in the production and manipulation of ultracold quantum gases.  This includes the production of Bose-Einstein condensates in electromagnetic traps \cite{Anderson1995,Davis1995,Courteille2001} or even in standing light fields \cite{Zoller1998,Jordens2008} that allow for an experimental implementation of Hubbard-type quantum models \cite{Greiner2002,Ott2008}.

These ultracold atomic gases have been studied in equilibrium situations for quite a while and with huge success. Nowadays, the focus shifts to investigating the nonequilibrium properties of such systems \cite{Mandel2003,Palzer2009,Hurtado2011,Ates2012,Chien2012a,Chien2012b,Wacker2013,Chien2013,Salger2013,Ventra2013}.

Recently, there have been the first real transport experiments with ultracold atoms \cite{Krinner2013,Brantut2012,Brantut2013}. In these setups, the transport processes are driven by at least two reservoirs which are initialized in different equilibrium states and attached to the system of interest. Accompanying these experimental advances, there has been also theoretical research involving atomic reservoirs coupled to, \eg, a lattice system \cite{Bruderer2012}, a potential trap \cite{Gutman2012}, or even quantum dot systems \cite{Ivanov2013}.

Since the ultracold atoms are trapped in a high-vacuum chamber, the gas is well separated from its environment such that no additional particle reservoir is present. Therefore, in these experiments the particle density is constant and the chemical potential can not be treated as free accessible parameter but becomes a function of the temperature and particle number in the trap. This strongly motivates us to investigate the influence of such a temperature- and particle-density-dependent chemical potential on the transport properties of a two-terminal open quantum system setup.

This analysis is especially interesting for systems that undergo quantum phase transitions. For example, transport through such systems contains information about such transitions even in extreme nonequilibrium setups \cite{Malte2012}. It should be noted that such transitions only occur in the thermodynamic limit (infinite system size), which suggests to investigate the role of criticality within the reservoirs. In fact, criticality is normally associated with a characteristic change in the chemical potential. Hence, it is crucial to describe the chemical potential in dependence of the thermodynamic state variables of the gas in order to correctly describe these critical phenomena. 

Within this paper, we particularly investigate the difference between the transport of fermionic and bosonic particles through such systems. Since an ideal Bose gas shows a quantum phase transition from a normal phase to a Bose-Einstein condensate in thermodynamic equilibrium, we expect to find signatures of this critical reservoir behavior in the transport coefficients.

In \secref{S:ThoereticalFramework}, we present the general theoretical framework that we use throughout this paper. We review the properties of ideal quantum gases in \secref{S:AtomicReservoirs}, derive the master-equation formalism with which we describe the transport through an open quantum system in \secref{S:MasterEquation}, and analyze the steady-state entropy production in \secref{S:EntropyProduction}. Finally, we apply a linear response theory to extract the linear transport coefficients in \secref{S:TransportCoefficients}. Subsequently, we apply this formalism to ideal Fermi gases in \secref{S:FermiGas} and to ideal Bose gases in \secref{S:BoseGas}, respectively, and summarize our results in \secref{S:Summary}.
%
%
\section{Theoretical Framework}\label{S:ThoereticalFramework}
%
%
The main difference of our setup compared to the usual schemes to describe transport through nanostructures lies in the utilization of massive ultracold atoms with a temperature- and density-dependent chemical potential. Consequently, the reservoirs are modeled as ideal quantum gases in the thermodynamic limit. 

Due to their statistical properties, bosonic reservoirs are fundamentally different from their electronic counterpart. In particular, the bosonic reservoirs undergo a quantum phase transition from a normal phase to a Bose-Einstein condensate \cite{Pitaevskii,Pethick}. Therefore, criticality is induced in the transport setup via the reservoirs. This critical behavior of the reservoirs is discussed in more detail in the following.
%
%
\subsection{Atomic Gases at Constant Density}\label{S:AtomicReservoirs}
%
%
The atomic baths are modeled as ideal gases of massive particles with a bath Hamiltonian given by
\begin{equation}\label{E:BathHamiltonian}
	\hat{H}_{\rm{B}}^{( \alpha)}=\underset{k}{\sum}\omega_{\alpha,k}\,\hat{b}_{\alpha,k}^{\dagger}\,\hat{b}_{\alpha,k},
\end{equation}
with operators $\hat{b}_{\alpha,k}^{\dagger}$ and $\hat{b}_{\alpha,k}$ creating and annihilating a particle with momentum $k$ and energy $\omega_{\alpha,k} = k^2/\rklamm{2 m_\alpha}$ in reservoir $\alpha$. Note that we use natural units throughout this paper, \ie,  $k_B=\hbar=1$. In the weak coupling limit the reservoirs enter additively, such that it suffices to consider here just a single reservoir. Therefore, we will drop the reservoir index $\alpha$ in this section. 

The mean occupation of the $k$-th energy level of an ideal quantum gas is given by $\bar n (\omega_k) = 1/\eklamm{e^{\beta(\omega_k -\mu)} - \xi}$, where $\xi=+1$ corresponds to a Bose gas and $\xi=-1$ corresponds to a Fermi gas. Here, we have introduced the inverse temperature $\beta=1/T$ and the chemical potential $\mu$. Since the lowest energy level of a free quantum gas is given by $\omega_0 =0 $ the positivity of the mean occupation demands that for bosons, in contrast to the electronic case, the chemical potential is restricted to negative values $-\infty < \mu \le 0$. 

The exact value of the chemical potential in the grand canonical ensemble is determined by the condition that the mean total number of particles
\equ{\bar N =\sum_k \bar n(\omega_k) = \sum_k \frac{1}{e^{\beta(\omega_k -\mu)} - \xi},}
is constant. We assume that the gas is confined in a three-dimensional ($3$D) cuboid of volume $V$ with periodic boundary conditions. In the thermodynamic limit where $\bar N \rightarrow \infty$, $V \rightarrow \infty$ with $n = \bar N / V = $const, the summation is replaced by an integral
$\frac{1}{\rklamm{2 \pi}^3} \sum_k \rightarrow \bint{0}{\infty}{\omega} g(\omega)$
with the density of states for an ideal, non-degenerate quantum gas given by
$g(\omega) = 2 \pi V g_s/\rklamm{2 \pi}^3 \rklamm{2 m}^{3/2} \omega^{1/2}$.
Here, $g_s=(2 S + 1)$ is the spin degeneracy coefficient. 

However, when replacing the sum by an integral, we need to take extra care of the ground-state occupation in the bosonic case, since in the regime where $-\beta \mu \ll 1$ it can be macroscopically occupied. This phenomenon does not occur in a Fermi gas due to the Pauli principle and is known as Bose-Einstein condensation \cite{Pitaevskii}. 

Keeping this in mind the mean total particle density is given by
\begin{align}
	n &= g_s\frac{\xi}{\lambda_T^3}{\rm{Li}}_{3/2}(\xi z)+n _0(\xi),\\
	n_0(\xi)&=
	\begin{cases}
	\frac{g_s}{V}\frac{z}{1-z}&:\xi=+1,\\
	0&:\xi=-1,
	\end{cases} \label{E:ParticleNumberLambda}
\end{align}
where we introduced the thermal de Broglie wavelength $\lambda_T = \sqrt{2 \pi / (m T)}$, the fugacity $z=e^{\beta \mu}$, and the poly-logarithm ${\rm{Li}}_s (z)= \sum_{k=1}^\infty z^k /k^s$ \cite{Arfken}. The explicit ground-state contribution $n_0(\xi)$ is only present in Bose gases. This equation implicitly defines the chemical potential $\mu= \mu(T,n)$ as a function of temperature and mean particle density.
%
\begin{figure}[t]
 \centering
 \includegraphics[width=\columnwidth]{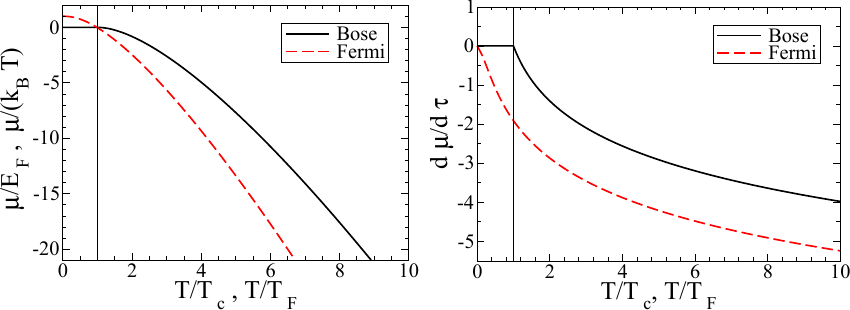}
 \caption{(Color online) Chemical potential (\textit{left}) and its derivative (\textit{right}) with respect to the normalized temperature for the ideal Bose (solid line) and the ideal Fermi gas (dashed line).}\label{F:ChemicalPotential}
\end{figure}

Due to the complexity of the poly-logarithm, the chemical potential can not be determined analytically and one has to use numerical methods. The results for the chemical potential of a Fermi and a Bose gas are depicted in \figref{F:ChemicalPotential}. Fortunately, all other quantities of interest can be expressed in terms of the chemical potential, leaving it to be the only numerical problem. Of special interest for the further calculations are the first derivatives of the chemical potential in the thermal phase which can be calculated from \equref{E:ParticleNumberLambda} to 
\equ{
    \left.\frac{\partial \mu }{\partial T}\right|_n =\frac{\mu }{T}-\frac{3 \text{Li}_{3/2}(\xi  z)}{2 \text{Li}_{1/2}(\xi  z)},\;\;
    \left.\frac{\partial \mu }{\partial n}\right|_T =\frac{T}{n}\frac{\text{Li}_{3/2}(\xi  z)}{\text{Li}_{1/2}(\xi  z)}.
}

In a real experiment with cold atoms, the chemical potential can not be tuned directly by applying an external voltage as usually considered for electronic transport. Instead one can introduce a thermal or density bias which causes a bias in the chemical potentials of the reservoirs. From the left panel in \figref{F:ChemicalPotential} we can see that applying a positive temperature bias at constant density where $T_1 > T_2$ results in an opposite chemical potential bias $\mu_2 > \mu_1$. The same effect occurs for a density bias at constant temperature. 

Furthermore, if the particle density is kept constant, one finds characteristic temperatures for the respective ideal quantum gases. For a Fermi gas, one defines the Fermi temperature $T_{\rm{F}}$ that relates to the Fermi energy $E_{\rm{F}}= T_{\rm{F}}$ which equals the chemical potential at absolute zero, \ie, $\mu(T=0K)=E_{\rm{F}}$.

For a Bose gas one finds a critical temperature $T_{\rm{c}}$ where the chemical potential vanishes and all particles start to condense in the same ground state. Thus, for temperatures below the critical point the Bose gas is in a mixed phase consisting of a normal thermal phase and a condensate fraction. When the temperature is absolute zero all particles occupy the ground state and the gas forms a pure Bose-Einstein condensate. 
These characteristic temperatures are defined by
\begin{align}
	T_{\rm{F}} = \frac{1}{2 m}\rklamm{ \frac{6 \pi^2 n}{g_s}}^{2/3}, \;
	T_{\rm{c}} = \frac{2 \pi}{m} \rklamm{ \frac{n}{g_s \zeta(3/2)} }^{2/3},
\end{align}
where $\zeta(s)$ is the Riemann zeta function. 

Analogous to the situation with constant density, we can also consider the case that the temperature is constant. This allows us to define a critical density $n_c$ for bosons and a Fermi-density $n_{\rm{F}}$ for fermions according to
\begin{align}
	n_c = g_s\frac{\zeta(3/2)}{\lambda_T^3}, \;
	n_{\rm{F}}= \frac{4 g_s}{3\sqrt{\pi} \lambda_T^3}.
\end{align}

In both cases, the chemical potential can be treated as a function of a single dimensionless variable $\tau = T/T_{\rm{c}}$, $\tau = T/T_{\rm{F}}$ or $\nu = n/n_{\rm{c}}$, $\nu = n/n_{\rm{F}}$ for ideal Bose and Fermi gases, respectively.

\begin{figure}[t]
 \centering
 \includegraphics[width=\columnwidth]{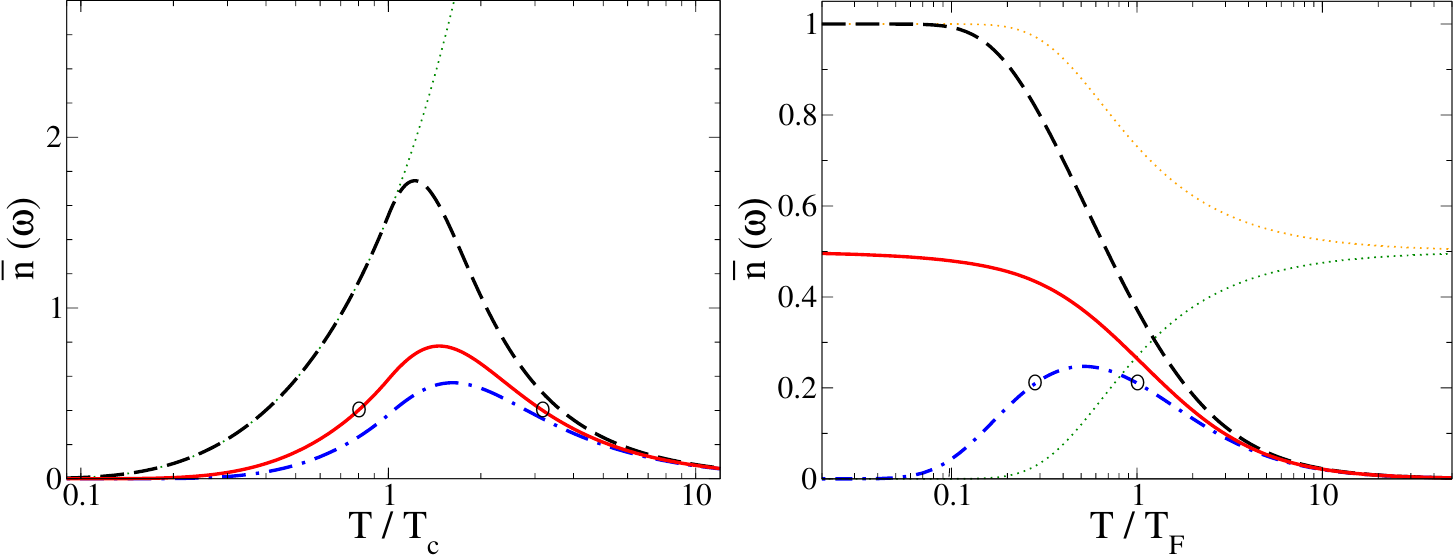}
 \caption{(Color online) Mean occupation for ideal quantum gases with temperature- and density-dependent chemical potential.  \textit{Left}: For an ideal Bose gas at $\omega=0.5 T_c$ (dashed line), $\omega=T_c$ (solid line) and $\omega=1.3 T_c$ (dotted-dashed line).  Using a constant chemical potential , \eg, $\omega-\mu=0.5 T_c$ (thin, dotted line) the occupation increases exponentially with the temperature. \textit{Right}: For an ideal Fermi gas at $\omega=0.5 T_{\rm{F}}$ (dashed line), $\omega=T_{\rm{F}}$ (solid line) and $\omega=1.3 T_{\rm{F}}$ (dotted-dashed line). The fermionic occupation for a constant chemical potential approaches $1/2$ for large temperature. This value is approached either from above if $\omega-\mu<0$, \eg, $\omega-\mu=-T_F$ (orange, dotted line) or from below if $\omega-\mu>0$, \eg, $\omega-\mu=T_F$ (green, dotted line). The circles indicate a set of temperatures with the same occupation for a given transition frequency [compare with Figs.\ (\ref{F:FermiCurrent}) and (\ref{F:
BoseCurrent})].}\label{F:MeanOccupation}
\end{figure}

Inserting the temperature- and density-dependent chemical potential into the definition of the  bosonic and fermionic mean occupations, we find a temperature dependence as shown in \figref{F:MeanOccupation}. 
In the bosonic case, we see that the mean occupation of a given energy level is peaked around a corresponding characteristic temperature. For low temperatures close to zero, the particles occupy lower energy levels and, thus, the mean occupation of the considered energy level decreases. Analogously, for high temperatures the particles are excited to higher energy levels and the mean occupation of the considered energy level decreases. 

In the case of ideal Fermi gases we find a similar behavior if the considered energy level lies above the Fermi energy (dotted-dashed line). However, the situation changes if one considers the occupation of an energy level below the Fermi energy (dashed line). Here, the occupation becomes constant if the temperature is decreased due to the Pauli principle. Contrary, in the case of an independent chemical potential, the bosonic mean occupation increases with increasing temperature (dotted line) and the fermionic mean occupation approaches the value $1/2$ as the temperature is increased (dotted line).

We therefore find that the temperature-dependent chemical potential strongly affects the high-temperature behavior of the mean occupations.
%
%
\subsection{Transport Master Equation}\label{S:MasterEquation}
%
\begin{figure}[t]
 \centering
 \includegraphics[width=.95\columnwidth]{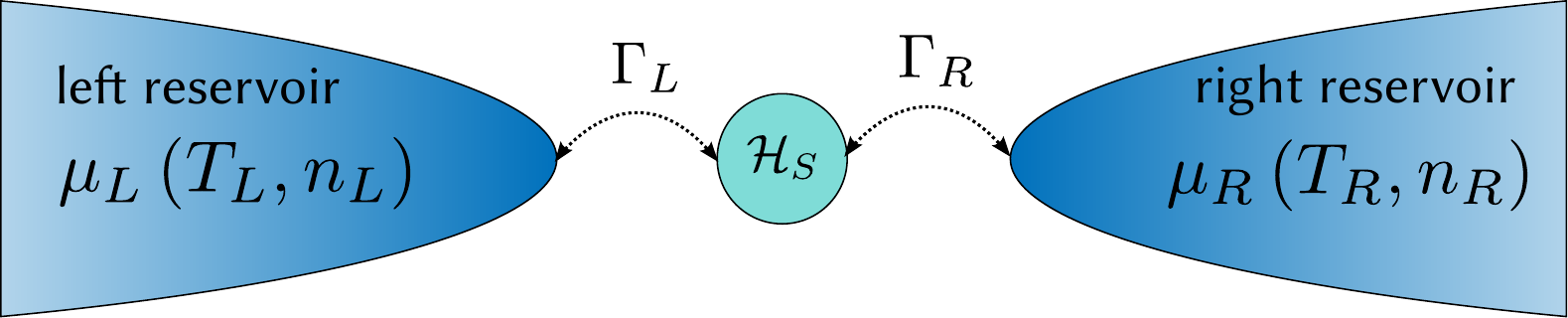}
 \caption{(Color online) General two terminal transport scheme with left and right reservoir weakly coupled to a few-level quantum system. The reservoirs $\alpha \in \sklamm{ L , R }$ are in thermal equilibrium and characterized by a chemical potential $\mu_\alpha(T_\alpha,n_\alpha)$ that depends on the respective temperature $T_\alpha$ and particle density $n_\alpha$. The system dynamics is governed by the Hamiltonian $\mathcal{H}_S$ and the weak system-bath coupling is mediated by tunneling rates $\Gamma_\alpha$. }\label{F:GeneralTransportScheme}
\end{figure}
%
We investigate a transport setup as depicted in \figref{F:GeneralTransportScheme} with two reservoirs in thermal equilibrium described by $T$ and $n$. These reservoirs coupled to the transport system are denoted by the labels $L$ and $R$.  We assume that the system-bath coupling is weak such that we can use the Born-Markov-Secular approximation (BMS) \cite{Breuer}. Starting from the von-Neumann equation this formalism allows to extract a quantum master equation that for non-degenerate system-energy eigenvalues assumes the form of a rate equation for the reduced system density matrix $\rho$ in the system-energy eigenbasis \cite{Schaller2008}. 

For sequential particle tunneling we can uniquely identify the jump terms in the master equation which enables one to convert it into a conditional master equation. This master equation is conditioned on the number $n$ of particles tunneled via one reservoir into or out of the system and the amount of energy $E$ transfered from this reservoir into the system. Due to conservation laws  we just need to consider one transport channel. Therefore, without loss of generality we will focus on the left reservoir only. The corresponding conditional master equation reads as
\begin{align}\label{E:BMSrateEquation}
 \dot\rho^{(E)}_n=\mathcal{L}_{0}\rho^{(E)}_n+\sum_\omega \rklamm{ \mathcal{L}^{+}_\omega\rho^{(E-\omega)}_{n-1}+\mathcal{L}^{-}_\omega\rho^{(E+\omega)}_{n+1}}.
\end{align}
Here, the super-operators $\mathcal{L}_{0}$, $\mathcal{L}^{+}_\omega$ and $\mathcal{L}^{-}_\omega$ are acting on the reduced system density matrix with $\mathcal{L}_{0}$ describing the internal dynamics and $\mathcal{L}^{+}_\omega$ and $\mathcal{L}^{-}_\omega$ describing jumps out of and into the system with transferred energy $\omega$, respectively. This particle-number and energy resolved master equation can also be established using virtual detectors as bookkeeping operators \cite{Gernot2009}.

Subsequently, we perform a Fourier transformation $\rho\left(\chi,\eta,t\right)=\sum_n \int dE\rho^{(E)}_n(t) \exp[\ii ( n \chi+ E \eta)]$ which introduces a particle counting field $\chi$ \cite{Levitov1996} and an energy counting field $\eta$ \cite{Nicolin2011} for the left reservoir. The resulting Liouville super operator for the left reservoir becomes a function of these counting fields and reads as
\begin{equation}\label{E:LiouvillianChiEta}
  \mathcal{L}\left(\chi,\eta\right)=\mathcal{L}_{0}+\sum_\omega \rklamm{\mathcal{L}^{+}_\omega e^{+\ii \chi+\ii \omega \eta }+\mathcal{L}^{-}_\omega e^{-\ii \chi - \ii \omega \eta}}.
\end{equation}

From this Liouvillian together with the normalization condition $\tr{\bar\rho}=1$, one can uniquely determine the steady-state reduced system density matrix $\bar\rho$ by solving the equation $0=\mathcal{L} (0,0)\,\bar\rho$. Subsequently, the steady-state particle current $J_N$ and energy current $J_E$ for the left reservoir are obtained by 
\equ{
 J_N=-\ii\,\rm{Tr}\left\lbrace\left. \partial_{\chi}\mathcal{L}(\chi,\eta)\right|_{{\chi}={\eta}=0}\,\bar\rho \right\rbrace,\label{E:StaticCurrentDefinition}\\
 J_E=-\ii\,\rm{Tr}\left\lbrace\left. \partial_{\eta}\mathcal{L}({\chi},{\eta})\right|_{{\chi}={\eta}=0}\,\bar\rho \right\rbrace.\label{E:StaticEnergyCurrentDefinition}
}
The particle and energy conservation implies the relations $J_N\equiv J_{N}^{(L)}=-J_{N}^{(R)}$ and $J_{E}\equiv J_{E}^{(L)}=-J_{E}^{(R)}$ for the currents measured at the left and right reservoirs, respectively. Note, that the currents $J_N$ and $J_E$ are defined as positive if the corresponding flow is directed from the left reservoir to the right reservoir.

Furthermore, we would like to point out that in real experiments with ultracold atoms the reservoirs contain a finite number of particles and energy only. Therefore, a flow through the system would on longer time scales (that may, however, be experimentally relevant) lead to an equilibration of the reservoirs. 

Assuming that the change of the temperature and density in each reservoir are both
linearly related to the respective heat current $J_Q$ and particle current $J_N$, \ie,
%
\equ{
     \dot T_\alpha(t) = \frac{1}{C_\alpha} J_Q^{(\alpha)}(t) \hspace{.3cm} \textrm{and} \hspace{.3cm}
     \dot n_\alpha(t) = \frac{1}{V_\alpha} J_N^{(\alpha)}(t)\,,
    }
%
where $C_{\alpha}$ denotes the heat capacity, one may establish a closed set of equations for $\rho(t)$, $T_\alpha(t)$, and $n_\alpha(t)$. These also determine the chemical potentials of the reservoirs $\mu_\alpha(t)$. For the case of temperature-independent chemical potentials, this method has already been applied \cite{Gernot2011a}.
%
%
\subsection{Entropy Production}\label{S:EntropyProduction}
%
%
For the considered transport setup we calculate the entropy production following the approach outlined in Ref.~\cite{Esposito2010a}. We rewrite the time derivative of the Shannon entropy of the system $\dot S=-\sum_i{\dot P_i \ln P_i}$ as a sum over an internal entropy production $\dot S_i$ and an entropy flow $\dot S_e$ from the system to the environment. If the system is in its steady state, $\bar{\dot S}=0$ vanishes and the internal entropy production of the system is given by the negative entropy flow to the environment $\bar{\dot S}_{i} = - \bar{\dot S}_e$. The steady-state entropy flow is given by 
\equ{
 \bar{\dot S}_e=\sum_{\alpha=L,R} \beta_\alpha \rklamm{J_{E}^{(\alpha)} - \mu_\alpha J_{N}^{(\alpha)}},
 }
where $J_E^{(\alpha)}$ and $J_N^{(\alpha)}$ denote the energy- and particle-currents from reservoir $\alpha$ into the system, respectively. Using the conservation of particle number and energy, we obtain the entropy production
\equ{
   \bar{\dot S}_i =&  J_{N} \Delta_{\mu\beta} - J_{E} \Delta_{\beta} \label{E:CommonEntropyProduction},
}
where we introduced the discrete affinities $\Delta_{\mu\beta}=\mu_{L}\beta_{L} - \mu_{R}\beta_{R}$ and $\Delta_\beta=\beta_{L} - \beta_{R}$. In the following, we assume a small thermal- and chemical potential bias between the reservoirs such that $T_L=T+\Delta_ T/2$, $T_R=T-\Delta_ T/2$ and $\mu_L=\mu+\Delta_\mu/2$, $\mu_R=\mu-\Delta_\mu/2$. Linearizing  \equref{E:CommonEntropyProduction} around the equilibrium, \ie, $\Delta_\mu=\Delta_T=0$, and collecting the terms associated to the affinity of the same intensive parameter yields the linear response entropy production as a function of the temperature and chemical potential
\equ{
      \bar{\dot S}_i(T,\mu) =&- J_{Q}  \Delta_\beta +  J_{N}  \beta \Delta_\mu. \label{E:UsualEntropyProduction}
}
Here, the quantity $J_{Q}=J_{E} - \mu J_{N}$ is the usual heat current. Furthermore, we assume that the chemical potential is not an independent parameter but a function $\mu(T,n)$ of the temperature and the particle density in the reservoir. Hence, the chemical potential bias in \equref{E:UsualEntropyProduction} has to be replaced by the linearized expression 
\equ{
  \Delta_\mu = \left.\frac{\partial \mu}{\partial n}\right|_T \Delta_n + \left.\frac{\partial \mu}{\partial \beta}\right|_n \Delta_\beta, \label{E:LinearizedChemicalPotentialBias}
  }
and we find the linear response entropy production as a function of the temperature and particle density
\equ{
       \bar{\dot S}_i(T,n) =& - \tilde{J}_{Q} \Delta_\beta  +  {J}_{N} \beta \frac{\partial \mu}{\partial n}\Delta_ n.\label{E:EntropyProduction}
}
Here, we introduced the generalized heat current
\equ{\tilde{J}_{Q} ={J}_{E}- \rklamm{\mu + \beta \frac{\partial \mu}{\partial \beta}} {J}_{N}, }
which corresponds to the conventional heat current $J_Q$ with a modified chemical potential. This modified chemical potential correctly describes the classical limit: For high temperatures, it assumes the value $\rm{lim}_{T\rightarrow\infty} \rklamm{\mu + \beta \partial \mu/\partial \beta}=3/2 k_B T$, which is the classical amount of heat per particle in three dimensions.

Thus, we find that the constraint of a constant particle density in the reservoirs leads to a modification of the chemical potential in the heat current and the emergence of a density driven particle current. Both of these contributions arise from the linearized affinity \eqref{E:LinearizedChemicalPotentialBias}. They can be interpreted as the work one needs to perform on a particle which is traveling through the transport setup, in order to overcome the chemical potential bias caused by the thermal or density bias, respectively. 
%
%
\subsection{Onsager Theorem}\label{S:TransportCoefficients}
%
%
It is well known that the Onsager theorem \cite{Onsager1931a,Onsager1931b} is very useful for describing linear, purely resistive systems. This theorem has been analyzed and proven to be also valid for open quantum systems \cite{Jaksic2002,Saito2008}. In particular, the Onsager theorem holds for open quantum systems which can be described by Markovian master equations \cite{Casimir1945,Moreauon1975,Hanggi1982}.  These are the quantum mechanical analog to purely resistive classical systems, \ie, systems without memory. Within this section we demonstrate the validity of this theorem and extract the linear response transport coefficients.

In order to appropriately describe an irreversible transport process one rewrites the entropy production as a sum $\bar{\dot S}_i =\sum_{j=1}^2 \mathcal{J}_j \mathcal{F}_j$ over generalized fluxes $\mathcal{J}_j$ and their corresponding affinities $\mathcal{F}_j$ \cite{Callen}. The linear response entropy production and, hence, the fluxes and affinities, will be different depending on whether we assume a constant particle density or not. To compare these two situations, we first consider the case without density constraint, \ie, for an independent chemical potential, and afterwards analyze the case with constant particle density.
\subsubsection{Independent chemical potential}\label{S:OnsagerIndependent}
First, we review the usual electronic transport approach with an independent chemical potential. Here, the entropy production is given by \equref{E:UsualEntropyProduction} and, thus, the generalized currents are given by
\equ{
     \mathcal{J}_1=  -{J}_{Q},\hspace{.5cm} \mathcal{J}_2= {J}_{N}, 
}
with the corresponding affinities
\equ{
    \mathcal{F}_1 = \Delta_\beta, \hspace{.5cm} \mathcal{F}_2 = \beta \Delta_ \mu.
}
Linearizing these currents with respect to their respective affinities around the equilibrium ($\Delta_\beta=0$, $\Delta_\mu=0$) yields an Onsager system in the form
\begin{equation}  
    \begin{pmatrix}
       - {J}_{Q}\\  {J}_{N} 
    \end{pmatrix} 
    = 
    \begin{pmatrix}
       L_{11} &  L_{12} \\   L_{21} &  L_{22}  
    \end{pmatrix}
    \begin{pmatrix}
      \Delta_\beta \\ \beta\Delta_\mu
    \end{pmatrix}
    \equiv
    {\boldsymbol{M}}
    \begin{pmatrix}
      \Delta_\beta \\ \beta\Delta_\mu
    \end{pmatrix},
    \label{E:UsualOnsagerSystem}
\end{equation}
where the entries of the Onsager matrix ${\boldsymbol{M}}$ with constant chemical potential are defined as derivatives evaluated at the equilibrium values, \ie, $L_{ij} = \rklamm{\partial \mathcal{J}_i/\partial \mathcal{F}_j}_0$. These so-called kinetic coefficients fulfill the Onsager reciprocal relation $L_{12}= L_{21}$ which is related to the time-reversal symmetry of physical laws \cite{Callen}. Furthermore, the Onsager matrix is positive definite which guarantees the positivity of the entropy production in accordance with the second law of thermodynamics.

From the Onsager system \eqref{E:UsualOnsagerSystem}, one can subsequently extract the linear transport relations for different setups. If no thermal bias is present, \ie, $\Delta_\beta=-1/T^2 \Delta_T=0$, one finds Ohm's law $J_N = \sigma \Delta_\mu$ with the electronic conductance $\sigma=L_{22}/T$. Similarly, one finds Fourier's law $J_Q = -\kappa \Delta_T$ for a thermocouple under the constraint of vanishing particle current $J_N=0$. This defines the linear heat conductance $\kappa=D/(T^2 L_{22})$ where $D=\rm{det}({\boldsymbol{M}})$ is the determinant of the Onsager matrix \eqref{E:UsualOnsagerSystem}. Additionally, such a system can produce a potential bias $\Delta_\mu = \Sigma \Delta_T$ as a response to a thermal bias at vanishing particle current. This so-called Seebeck effect is characterized by the thermopower $\Sigma=L_{21}/(T L_{22})$. The reverse process where a thermal bias is created by applying a bias in the chemical potentials is known as Peltier effect which is characterized by the Peltier 
coefficient $\Pi = T \Sigma$. The efficiency of these processes can be characterized by the dimensionless figure-of-merit $ZT=\Sigma^2/L$ with the Lorenz number $L=\kappa/(T \sigma)$ defined by the Wiedemann-Franz law \cite{Wiedemann}.
\subsubsection{Dependent chemical potential}\label{S:OnsagerDependent}
In analogy to the discussion in the previous paragraph, we now focus on the situation where $T$ and $n$ are independent variables. When the temperature and particle-density in the reservoirs are held at constant differences, the entropy production is given by \equref{E:EntropyProduction} with generalized currents
\equ{
     \mathcal{J}_1=  -{\tilde J}_{Q},\hspace{.5cm} \mathcal{J}_2= {J}_{N}, 
}
and their respective affinities \cite{Affinity}
\equ{
    \mathcal{F}_1 = \Delta_\beta, \hspace{.5cm} \mathcal{F}_2 = \beta \frac{\partial\mu}{\partial n}\Delta_ n.
}
The corresponding linearized Onsager system reads as
\begin{equation}  
    \begin{pmatrix} - \tilde{J}_{Q}\\  {J}_{N} \end{pmatrix} 
    = 
    \begin{pmatrix}
      \tilde L_{11} & \tilde L_{12} \\  \tilde L_{21} & \tilde L_{22}  
    \end{pmatrix}
    \begin{pmatrix}
      \Delta_\beta \\ \beta \frac{\partial\mu}{\partial n}\Delta_n
    \end{pmatrix}
    \equiv
    {\boldsymbol{\tilde M}}
    \begin{pmatrix}
      \Delta_\beta \\  \beta \frac{\partial\mu}{\partial n}\Delta_n
    \end{pmatrix},
    \label{E:OnsagerSystem}
\end{equation}
where the kinetic coefficients $\tilde{ L}_{ij}=\rklamm{\partial \mathcal J_i/\partial \mathcal F_j}_0$ are now functionals of the chemical potential $\mu(T,n)$. 

Due to the linearity of the system of equations, we can find a linear mapping which transforms the Onsager matrices in Eqs.\ (\ref{E:UsualOnsagerSystem}) and (\ref{E:OnsagerSystem}) into each other (see Appendix \ref{A:OnsagerMatrices}). Hence, we can rewrite the matrix $\boldsymbol{\tilde M}$ using the kinetic coefficients defined in \equref{E:UsualOnsagerSystem} which now become functionals of the temperature- and density-dependent chemical potential, \ie, $L_{ij}(\mu)\rightarrow L_{ij}[\mu(T,n))]$. This yields
\begin{align}
  {\boldsymbol{\tilde M}}
  =
  \left(
\begin{array}{cc}
 L_{11}+\beta \frac{\partial \mu  }{\partial \beta }\left(2 L_{12}+\frac{\partial \mu  }{\partial \beta }L_{22}\right) & L_{12}+\beta \frac{\partial \mu  }{\partial \beta }L_{22} \\
 L_{21}+\beta \frac{\partial \mu  }{\partial \beta }L_{22} & L_{22}
\end{array}
\right),\label{E:OnsagerCorrespondence}
\end{align}
where the Onsager reciprocal relation is preserved, \ie, $\tilde L_{12}= \tilde L_{21}$. 
From the above equation, we derive linear transport coefficients analogous to the electronic case.

We see that in the absence of a thermal bias $\Delta_ T=0$, the particle current ${J}_{N}=\beta \tilde L_{22} \frac{\partial\mu}{\partial n}\Delta_ n$ becomes proportional to the applied density bias. This yields an equation similar to Ohm's law ${J}_{N} = \tilde\sigma  \frac{\partial\mu}{\partial n}\Delta_ n$ with an isothermal matter conductance $\tilde\sigma$ given by
\equ{
    \tilde\sigma \equiv& \frac{{J}_{N}}{ \frac{\partial\mu}{\partial n}\Delta_ n} = \frac{\tilde L_{22}}{T}= \sigma[\mu(T,n)] ,\;\textrm{for}\;\;\Delta_T = 0.\label{E:MatterConductance}
}

In a similar way to the matter conductance, we can extract the analog of the thermal conductance $\tilde\kappa$ from the modified Fourier's law $\tilde{J}_{Q}=-\tilde\kappa \Delta_ T$ under the constraint ${J}_{N} =0$ which yields 
\equ{ 
  \tilde\kappa \equiv  -\frac{\tilde{J}_{Q}}{\Delta_ T} =\frac{\tilde D}{T^2 \tilde L_{22}} = \kappa[\mu(T,n)].\label{E:ThermoConductance}
}
Here, $\tilde D={\rm{det}}(\boldsymbol{\tilde M})$ is the determinant of the Onsager matrix $\boldsymbol{\tilde M}$.
Note that this transport coefficient vanishes if the determinant is zero. In general, this happens in the so-called tight-coupling limit  where the energy current $J_E$ becomes proportional to the particle current $J_N$ and, hence, the generalized fluxes $\mathcal{J}_1$ and $\mathcal{J}_2$ are not independent of each other \cite{Broeck2005,Broeck2012}.

Furthermore, we find that a vanishing particle current ${J}_{N}=0$ for finite thermal and density bias implies $\beta \frac{\partial\mu}{\partial n} \Delta_n  =  \tilde L_{21}/(T^2 \tilde L_{22}) \Delta_ T$. Therefore, such a thermodynamic device produces a density-induced chemical potential bias as a response to a thermal bias. This allows us to define $\tilde \Sigma$ analogous to the thermopower by
\equ{\tilde\Sigma \equiv \frac{\frac{\partial\mu}{\partial n}\Delta_n}{\Delta_T}= \frac{\tilde L_{21}}{T \tilde L_{22}}  = \Sigma[\mu(T,n)] + \beta^2 \frac{\partial\mu}{\partial\beta}.\label{E:Thermopower}}
This coefficient characterizes the linear density response to a temperature difference at vanishing particle current.
It is related to the analog of the Peltier coefficient $\tilde \Pi$ by the Thomson relation $\tilde \Pi = T \tilde \Sigma$. 
Using these transport coefficients, we can calculate the dimensionless figure-of-merit $\tilde {ZT}$  \cite{Apostol2008} which characterizes the efficiency of the thermodynamic device. It is given by 
\equ{
 \tilde {ZT} \equiv \frac{\tilde\Sigma^2}{\tilde L}  = {ZT}[\mu(T,n)] + \frac{\beta^2 \frac{\partial\mu }{ \partial \beta }\left(\beta^2 \frac{\partial\mu }{ \partial \beta }  + 2 \Sigma[\mu(T,n)] \right)}{L[\mu(T,n)]},\label{E:FigureOfMerit}
}
with the modified Lorenz number $\tilde L=\tilde \kappa/(\tilde\sigma T)$. From the definitions (\ref{E:MatterConductance}) - (\ref{E:FigureOfMerit}) we see that for an independent chemical potential, where the derivative with respect to temperature vanishes, \ie, $\partial\mu/\partial\beta=0$, we recover the usual linear response transport coefficients. 
%
%
\section{Ideal Fermi Gases}\label{S:FermiGas}
%
%
\begin{figure}[t]
 \includegraphics[width=0.7\columnwidth]{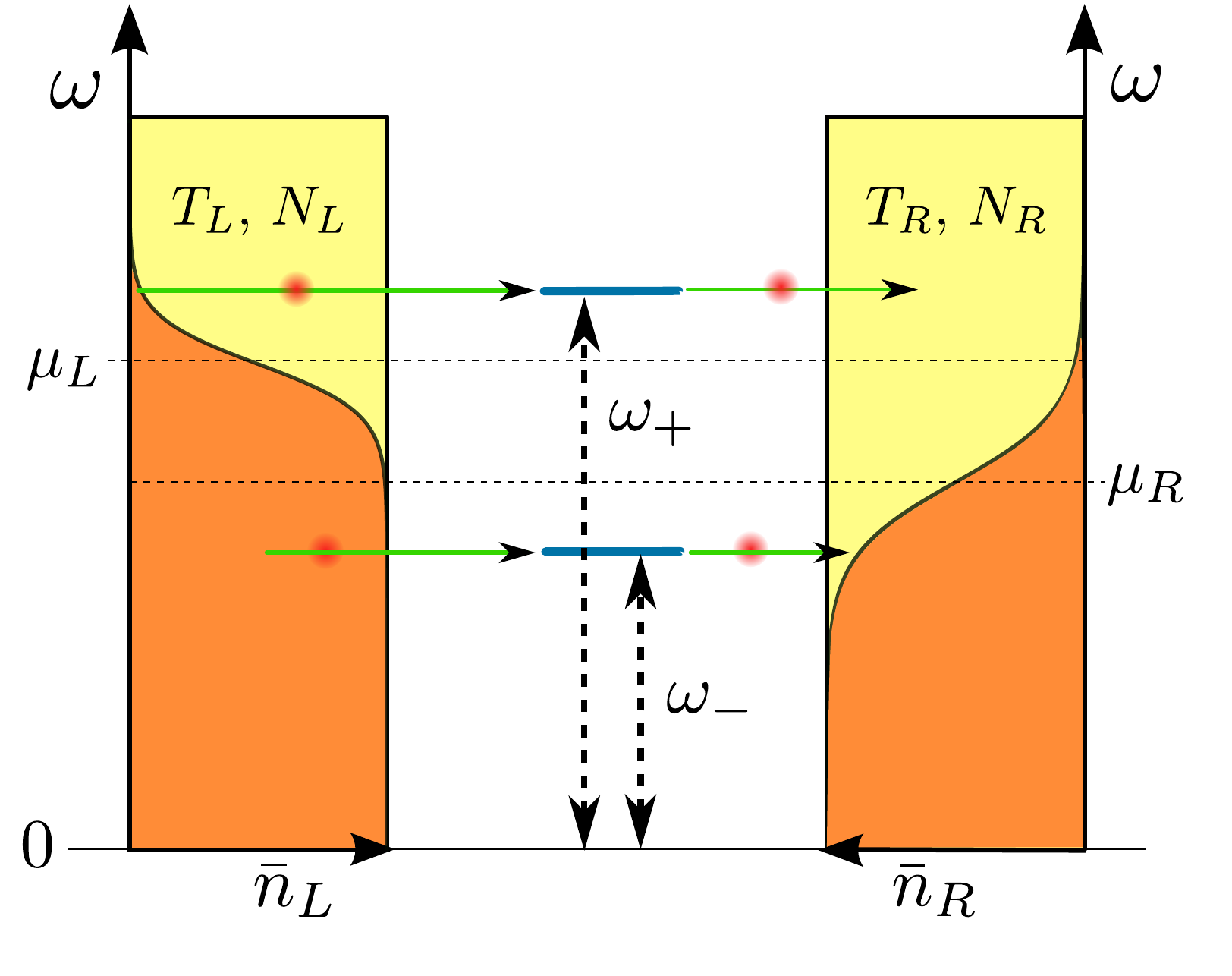}
 \caption{(Color online) Setup for fermionic particle transport. The atomic reservoirs $\alpha \in \sklamm{L,R}$ are in thermal equilibrium characterized by temperature $T_\alpha$ and chemical potential $\mu_\alpha=\mu(T_\alpha,N_\alpha)$ for fixed particle number $N_\alpha$. The mean occupation $\bar n_\alpha(\omega)$ of the transition energy $\omega$ is given by the Fermi-Dirac distribution. The system is composed of a double quantum dot in the Coulomb blockade regime with two transition energies $\omega_-=\varepsilon-g$ and $\omega_+=\varepsilon+g$ only.}
 \label{F:FermiModel}
\end{figure}
%
As a first example, we consider a fermionic system as shown in \figref{F:FermiModel} which is composed of a double quantum dot in the Coulomb blockade regime and two fermionic terminals. The system is described by the Hamiltonian
\begin{align}\label{E:SysHamiltonianFermi}
 \hat{\mathcal{H}}_{\rm{S}}^{\rm{fermi}} =&\, \varepsilon \rklamm{\opd{c}_{\rm{L}}\op{c}_{\rm{L}}+\opd{c}_{\rm{R}}\op{c}_{\rm{R}}} + g \rklamm{\opd{c}_{\rm{L}}\op{c}_{\rm{R}}+\opd{c}_{\rm{R}}\op{c}_{\rm{L}}} \notag\\ 
 &+ V \opd{c}_{\rm{L}}\op{c}_{\rm{L}}\opd{c}_{\rm{R}}\op{c}_{\rm{R}}. 
\end{align}
Here, the operators $\op{c}_\alpha$ and $\opd{c}_\alpha$ which obey the fermionic anti-commutation relation $\sklamm{ \op{c}_\alpha,\opd{c}_{\alpha}}=1$ annihilate and create a fermion particle in quantum dot $\alpha$, respectively. These two quantum dots are labeled by $L$ and $R$ and they are coupled via a coherent tunneling process with amplitude $g$. Each dot can be empty or occupied by a single particle increasing the system energy by $\varepsilon$. In the Coulomb blockade limit the Coulomb repulsion $V\gg \varepsilon, g$ is the dominating energy scale. Hence, the state corresponding to a doubly occupied double quantum dot does not take part in the long-time  dynamics and can be safely neglected. The remaining energy eigenstates of the system are the vacuum state $\ket{0}$ and the superposition states $\ket{-}=1/\sqrt{2}\rklamm{\ket{01}-\ket{10}}$ and $\ket{+}=1/\sqrt{2}\rklamm{\ket{01}+\ket{10}}$ with eigenvalues $\omega_0=0$, $\omega_-=\varepsilon-g$ and $\omega_+=\varepsilon+g$, respectively.

The system is coupled to the reservoirs by the system-bath interaction Hamiltonian
\begin{equation}\label{E:FermiSysBathHamiltonian}
	\hat{H}_{\rm{SB}}=\underset{\alpha,k}{\sum} \left( t_{\alpha,k} \, \hat{b}_{\alpha,k}^{\dagger}\,\hat{c}_\alpha + \rm{H.\,c.} \right),
\end{equation}
where the tunneling amplitude of a particle hopping from the reservoir $\alpha$ into the respective quantum dot or vice versa is proportional to $t_{\alpha,k}^{*}$ and $t_{\alpha,k}$, respectively. 
\subsection{Steady-State Current}\label{S:FermiSteadyStateCurrent}
\begin{figure}[t]
 \centering
 \includegraphics[width=\columnwidth]{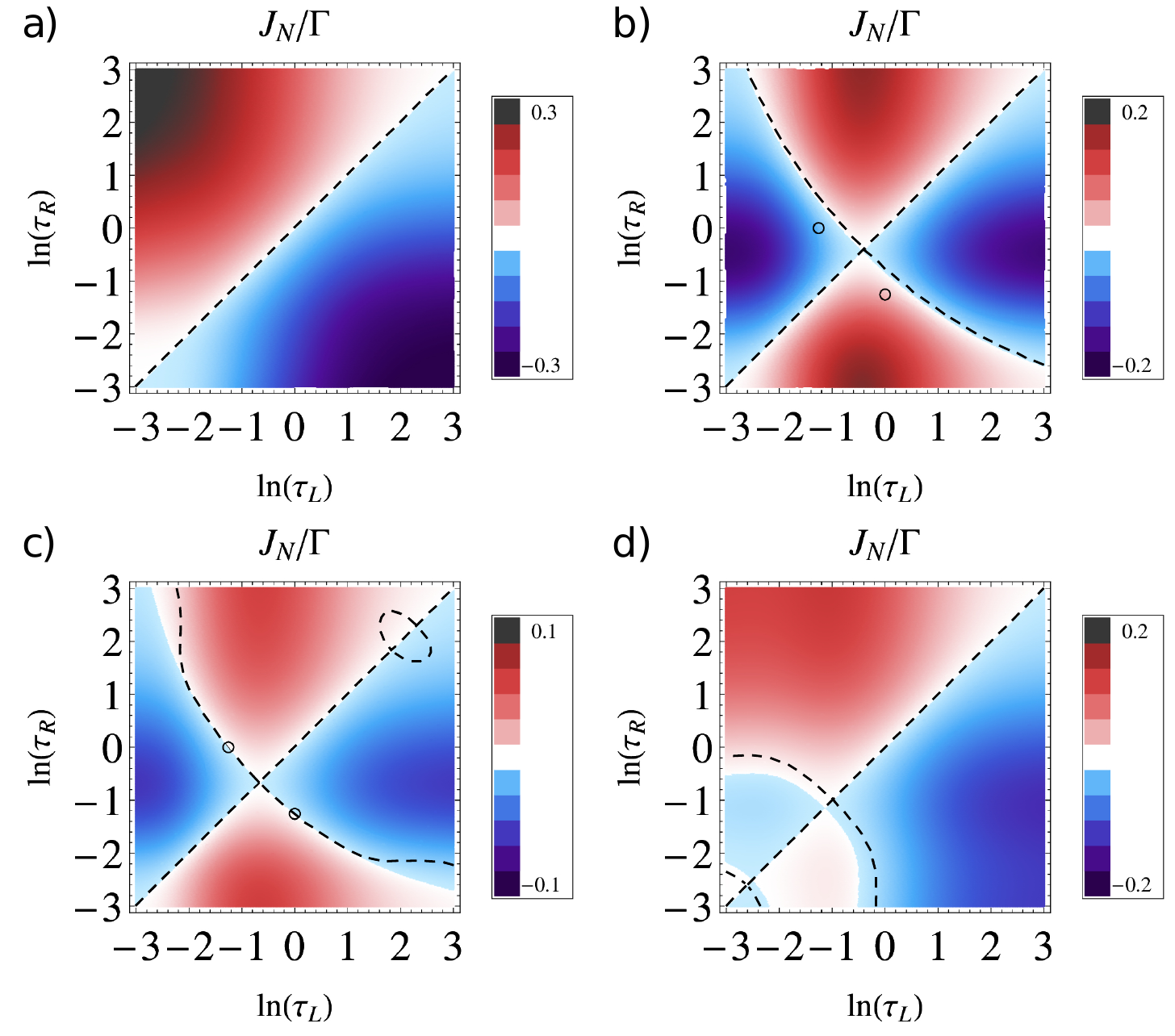}
 \caption{(Color online) Steady-state particle current of the fermionic system with different transition frequencies versus the dimensionless temperatures of the reservoirs at fixed density. In the plots, we set $\varepsilon=0.7 E_F$ (a), $\varepsilon= 1.5 E_F$ (b), and $\varepsilon=1.2 E_F$ (d) and used the same tunneling amplitude $g=0.2 E_F$. In plot (c), we used $\varepsilon=11.3 E_F$ and set the tunneling amplitude to $g=10 E_F$. For all plots, the rates are set to $\Gamma_L = \Gamma_R = \Gamma$. The circles in the plots (b) and (c) correspond to the set of temperatures marked in \figref{F:MeanOccupation}. The dashed curves indicate a vanishing of the corresponding energy current $J_E$.} \label{F:FermiCurrent}
\end{figure}

We start by calculating the steady-state particle and energy currents according to \equref{E:StaticCurrentDefinition}  (see Appendix \ref{A:FermionicLiouvillian}). This yields for the steady-state current measured at reservoir $\alpha$ the relations
\begin{align}
    J_N^{(\alpha)} =& \sum_{n \in \sklamm{+,-}} \mathcal{I}_n^{(\alpha)},\, J_E^{(\alpha)} = \sum_{n \in \sklamm{+,-}} \omega_n \mathcal{I}_n^{(\alpha)},
\end{align}
where we defined the abbreviation
\equ{ \mathcal{I}_n^{(\alpha)} =-\frac{\Gamma_\alpha}{2}\left\{ \bar{n}_\alpha(\omega_n)  \bar{\rho}_0 -\eklamm{1-\bar n_\alpha(\omega_n)} \bar\rho_{n}\right\},}
with the steady-state density vector $\bar\rho=\rklamm{\bar\rho_0,\bar\rho_-,\bar\rho_+}^{\rm{T}}$.
Since the complete expression is too long we state here the particle current in the limit of a single transition frequency only. This limit can be obtained by shifting the second transition energy to high values such that transport through this level is strongly suppressed. In consequence, we find a current involving the lowest transition energy only which reads as
\equ{
  \lim_{\omega_+\rightarrow\infty} {J}_{N} = \frac{\bar\Gamma}{2}\left[\bar n_{L}\left(\omega_-\right)-\bar n_R\left(\omega_-\right)\right],\label{E:FermiSingleFrequencyCurrent}
}
with the effective coupling rate $\bar\Gamma=\Gamma_L \Gamma_R/  \left(\Gamma _L+\Gamma _R\right)$.
Thus, the particle current through a system with transition energy $\omega_-$ is proportional to the difference of the mean occupations of the corresponding energy level in the reservoirs.
The particle current through a double-dot system with two transition frequencies is shown in \figref{F:FermiCurrent}.
%

We observe two different regimes reflecting the different behavior of the mean occupations for energies below and above the Fermi energy as shown in \figref{F:MeanOccupation}. If at least one transition energy lies below the Fermi energy, as shown in \figref{F:FermiCurrent}(a), we observe a finite steady-state current against the thermal bias. This is due to the fact that in the hotter reservoir the particles are excited to higher energy-states. Since the density is fixed, there are not enough particles to refill the depleted energy levels. The occupation in these levels decreases leading to a flow from the colder reservoir, where the energy levels are occupied, to the hotter reservoir. This behavior is a consequence of the mean occupation under the restraint of constant particle density.

If all transition energies are above the Fermi energy, as shown in Figs.\ \ref{F:FermiCurrent}(b) and \ref{F:FermiCurrent}(c), this behavior changes such that for a small thermal bias the steady-state current flows with the bias. However, if the thermal bias is increased above a critical value the current flows against the bias again. Moreover, there is always a finite steady-state current for an arbitrary high thermal bias although it is exponentially suppressed away from the optimal temperature. 

Taking a look at \equref{E:FermiSingleFrequencyCurrent} we see that the critical lines where the current vanishes are defined by the relation $\bar n_{L}\left(\omega_-\right)=\bar n_R\left(\omega_-\right)$. This is trivially fulfilled in equilibrium where $\Delta_\beta=\Delta_{n}=0$. Away from equilibrium we find that the particle current \eqref{E:FermiSingleFrequencyCurrent} only vanishes if the mean occupation of a given transition energy in the reservoirs takes on the same value for different temperatures. Comparing with the result presented in \figref{F:MeanOccupation} we can immediately deduce that this condition can only be satisfied for transition frequencies above the Fermi energy. In this case one always finds a set of two different temperatures for the left and right reservoirs where the current vanishes.

As an example, we indicated such a set of temperatures in \figref{F:MeanOccupation} and show the corresponding points in the current plots in Figs.\ \ref{F:FermiCurrent}(b) and \ref{F:FermiCurrent}(c).
In \figref{F:FermiCurrent}(c), we chose the dot energy $\varepsilon$ and the tunneling amplitude $g$ in such a way that the transition energy $\omega_+$ is shifted to high energies. Thus, the corresponding particle current can be approximately described by the single-level limit given in \equref{E:FermiSingleFrequencyCurrent}. 

Comparing the currents plotted in Figs.\ \ref{F:FermiCurrent}(b) and \ref{F:FermiCurrent}(c), we see that for a system with two transition frequencies [\figref{F:FermiCurrent}(b)] the line of vanishing particle current it shifted to higher temperatures compared to the effective single-level result [\figref{F:FermiCurrent}(c)]. This effect results from the additional transport channel which modifies the condition for a vanishing particle current. In fact, depending on the number and values of the transition frequencies in the system, there can also by more lines where the particle current vanishes [see panel \figref{F:FermiCurrent}(d)].

Finally, we note that the energy current $J_E$ in general vanishes (dashed lines) for different parameters than the particle current $J_N$. Thus, we can observe a finite energy-current even for a vanishing particle-current in a fermionic system with two transition frequencies. Moreover, in the upper right corner of \figref{F:FermiCurrent}(c) we even find a regime for high temperatures where the energy current flows against the particle current.
\subsection{Transport Coefficients}\label{S:FermiTransportCoefficients}
In this section we calculate the linear transport coefficients for the fermionic transport setup. For reasons of brevity, we use the wide-band limit with energy independent rates $\Gamma_\alpha(\omega)=\Gamma_\alpha$.
%
\begin{figure}[t]
 \centering
 \includegraphics[width=\columnwidth]{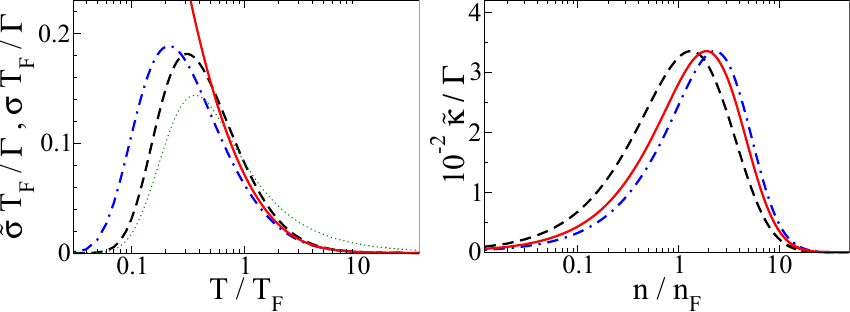}
 \caption{(Color online) Plot of the matter conductance $\tilde\sigma$, conductance $\sigma$ \textit{(left)} and the thermal conductance $\tilde\kappa$ \textit{(right)} for ideal Fermi gas reservoirs with different on-site energies $\varepsilon=0.7 E_F$ (dashed line), $\varepsilon=1.5 E_F$ (dotted-dashed line) and $\varepsilon=1.2 E_F$ (solid line) versus the normalized temperature and density, respectively. In both plots we assume equal tunneling rates $\Gamma_L=\Gamma_R=\Gamma$ and a constant coherent coupling strength of $g=0.2 E_F$. For the conductance with constant chemical potential we use $\varepsilon=1.2 E_F$, $g=0.2 E_F$ and additionally set $\mu=0.5 E_F$ (dotted line).}\label{F:FermiConductance}
\end{figure}

We plot some results for these transport coefficients in \figref{F:FermiConductance} and \figref{F:FermiSeeBeck}. In all these plots we analyze three different transport channel configurations. The first configuration corresponds to a system with transition energies below the Fermi energy of the reservoirs (dashed line). The second configuration corresponds to a system with transition energies above the Fermi energy of the reservoirs (dotted-dashed line). In the third configuration we analyze a system whose lowest transition energy is exactly equal to the Fermi energy of the reservoirs (solid line). The results in these situations are discussed in more detail within the following subsections. 
\subsubsection{Matter Conductance}\label{S:FermionicMatterConductance}
Calculating the matter conductance in the wide-band limit according to \equref{E:MatterConductance} yields the relation
\equ{
    \tilde\sigma =\frac{\bar\Gamma\left[1-\bar{n}\left(\omega _-\right)\right] \left[1-\bar{n}\left(\omega _+\right)\right] \left[\bar{n}\left(\omega _-\right)+ \bar{n}\left(\omega _+\right)\right]}{2 T \left[1-\bar{n}\left(\omega _-\right) \bar{n}\left(\omega _+\right)\right]}.\label{E:FermiDiffusion}
}
For $\bar n(\omega_+)=0$, \ie, in the limit of a single transition frequency $\omega_+ \rightarrow \infty$, this expression coincides with $\lim_{\omega_+\rightarrow\infty} \tilde\sigma=\bar\Gamma/\rklamm{8 T \rm{cosh}^2\eklamm{\rklamm{\omega_- -\mu}/(2 T)}}$, the well-known Coulomb blockade conductance peak for a single resonant level \cite{Beenakker1991}. Similarly, in the limit $g\rightarrow0$, \ie, $\omega_-\rightarrow\omega_+$ the matter conductance approaches a limiting value. However, at $g=0$ the energies $\omega_-=\omega_+=\varepsilon$ are degenerate and our rate equation approach can not be applied. Furthermore, the whole temperature dependence of the chemical potential enters implicitly via the mean occupations of the energy levels in the reservoirs in compliance with \equref{E:MatterConductance}. 

In the left panel of \figref{F:FermiConductance} we plot the matter conductance as a function of the normalized temperature for different on-site energies and a constant tunneling amplitude. 
For a configuration where both transition energies lie below the Fermi energy of the reservoirs (dashed line) we observe a maximal conductance at a specific temperature which basically depends on the frequency $\varepsilon$. Decreasing the temperature further diminishes the conductance as the respective energy levels in the reservoirs become occupied and the current decreases.
We observe a similar behavior for a configuration where both transition energies lie above the Fermi energy of the reservoirs (dotted-dashed line). In this situation the conductance vanishes for decreasing temperature due to the fact that the transition energy in the reservoirs gets exponentially depleted.

Finally, we show the result for a configuration where the lower transition energy equals the Fermi energy of the reservoirs (solid line). Only in this case we observe a non-vanishing conductance as the temperature approaches $0$. This is due to the fact that the Fermi energy level in the reservoirs is at most half filled whereas all other energy levels are either completely filled or empty. Therefore, particle transport is possible even for low temperatures. However, the observation that in this case the matter conductance $\tilde\sigma$ diverges like $\bar\Gamma/T$ is unphysical. This behavior stems from the fact that the Born-Markov-Secular master equation breaks down if $T\ll\bar\Gamma$. 
In all three situations the conductance vanishes for increasing temperature due to the reduction of the occupation of the energy levels in the reservoirs.

For comparison, we additionally plotted the conductance $\sigma$ for a constant chemical potential (dotted line). In this situation, we find qualitatively the same behavior as for the modified matter conductance. For low temperatures, the respective energy levels in the reservoirs are depleted and, hence, the conductance vanishes. Contrary for high temperatures they are equally filled which leads to a vanishing net current. Thus, the modified matter conductance $\tilde \sigma$ basically follows $\sigma$. However, the high-temperature behavior is changed due to the temperature dependence of the chemical potential.
\subsubsection{Heat Conductance}
From \equref{E:ThermoConductance} we find that the heat conductance in the wide-band limit is given by
\equ{
    \tilde\kappa = \frac{\bar n\left(\omega_{-}\right) \bar n\left(\omega_{+}\right) \left(\omega_{-}-\omega_{+}\right)^2}{T \left[\bar n\left(\omega_{-}\right)+ \bar n\left(\omega_{+}\right)\right]^2}\tilde\sigma.\label{E:FermiThermalConductance}
}
This expression has no explicit dependence on the chemical potential and, thus, formally corresponds to the thermal conductance for an independent chemical potential as shown in \secref{S:OnsagerDependent}. Furthermore, we immediately see that the above equation obeys the Wiedemann-Franz law, \ie, $\tilde\kappa = T \tilde L  \tilde \sigma$ with the dimensionless Lorenz number $\tilde L $. In the limit of a single transition-frequency, \ie, $\omega_+\rightarrow\infty$, the heat conductance $\tilde \kappa$ vanishes trivially because there is no pure heat flow through a single level without particle flow.

In the right panel of \figref{F:FermiConductance} we plot the modified thermal conductance versus the inverse normalized particle density for different on-site energies and different tunneling amplitudes according to \equref{E:FermiThermalConductance}. 
For all considered configurations, we observe qualitatively the same behavior. The heat transport is maximal at a characteristic density. This maximum is shifted to higher densities with increasing transition energies. For low densities, the thermal conductance vanishes as the reservoir energy levels become less occupied. For high densities, the heat conductance vanishes because the transition energies in both reservoirs become maximally occupied and the matter conductance vanishes. 
\subsubsection{Thermopower}
%
\begin{figure}[t]
 \centering
 \includegraphics[width=\columnwidth]{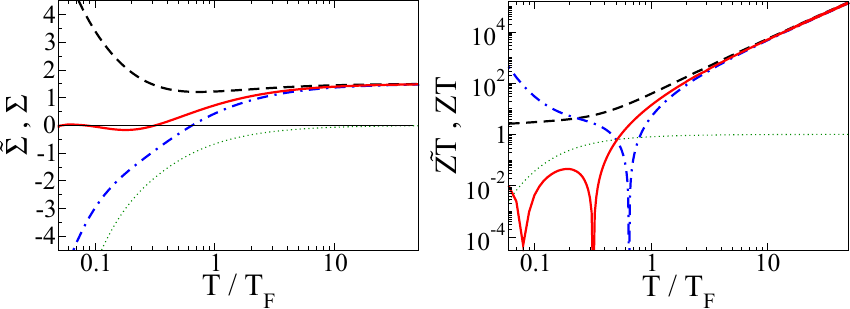}
 \caption{(Color online) Plot of the modified thermopower $\tilde\Sigma$ and conventional thermopower $\Sigma$ (\textit{left}) and the figure-of-merit for dependent and constant chemical potential (\textit{right}) versus the normalized temperature for an ideal Fermi gas and different symmetric on-site energies $\varepsilon=0.7 E_F$ (dashed line), $\varepsilon=1.2 E_F$ (solid line) and $\varepsilon=1.5 E_F$ (dotted-dashed line). For $\varepsilon=1.2 E_F$ and a constant chemical potential $\mu=0.5 E_F$ (thin, dotted line) the thermopower approaches $0$ for large T and the figure-of-merit takes on a constant value depending on the system transition frequencies. In both plots, we fixed the coherent tunneling amplitude at $g=0.2 E_F$. }\label{F:FermiSeeBeck}
\end{figure}
%
Next, we determine the analog to the thermopower defined in \equref{E:Thermopower}. For our setup we find the expression 
\equ{\tilde\Sigma = \frac{\braket{\phi_f(\omega_-,\omega_+)}}{T},}
where we defined the average energy
\equ{
  \braket{\phi_f(\omega_-,\omega_+)}=\frac{\bar{n}\left(\omega _-\right) \phi(\omega_-)+\bar{n}\left(\omega _+\right) \phi(\omega_+)}{\bar{n}\left(\omega _-\right)+\bar{n}\left(\omega _+\right)}.\label{E:MeanTransportedEnergy}
} 
Here, the expression
\equ{\phi(\omega_i)=\mu - T \partial\mu/\partial T - \omega_i,\label{E:SingleParticleHeat}}
describes the amount of energy one particle traveling through a transport system with only a single transition-energy $\omega_i$ takes from one reservoir to the other.
We can formally recover the result for the conventional thermopower with an independent chemical potential by setting  $\partial\mu/\partial T=0$.

In the left panel of \figref{F:FermiSeeBeck} we plot the temperature dependence of the modified thermopower $\tilde\Sigma$ for different transition energies in the case of Fermi reservoirs according to \equref{F:FermiSeeBeck}.
When the temperature decreases we observe different behavior for the modified thermopower depending on the transport system. If the lower transition energy is below the Fermi energy the modified thermopower remains positive for all temperatures. As the temperature approaches zero the average energy \eqref{E:MeanTransportedEnergy} approaches a constant positive value and, thus, the modified thermopower diverges like $1/T$. 

If the transition energies are above the Fermi energy we observe a similar behavior but the coefficient becomes negative since below some critical temperature the transition frequencies exceed the chemical potentials. 
Only for the case when the lower transition energy equals the Fermi energy the average energy and hence the modified thermopower vanishes at $T=0$. 

For high temperatures the average energy $\braket{\phi_f(\omega_-,\omega_+)}$ is dominated by the classical thermal energy contribution of $3/2 k_B T$ per particle. Therefore, we observe that the modified thermopower assumes a constant positive value of $3/2$ for high temperature independent of the respective transition energies. 

This behavior can not be predicted if the chemical potential is treated as an independent parameter. In this conventional approach the thermopower vanishes for high temperatures independent of the transition frequencies (dotted line). Furthermore, the conventional thermopower never changes its sign as a function of the temperature. In this situation the sign is fixed by the choice of the constant chemical potential.
\subsubsection{Figure-of-merit}
Finally, we analyze the figure-of-merit for the thermodynamic device. This coefficient is defined in \equref{E:FigureOfMerit} which relates to the efficiency of the device. For the considered fermionic setup the figure-of-merit reads as
\equ{
    \tilde{ZT} =\frac{\left[\bar{n}\left(\omega _-\right)+\bar{n}\left(\omega _+\right)\right]^2 \braket{\phi_f(\omega_-,\omega_+)}^2}{\bar{n}\left(\omega _-\right)\bar{n}\left(\omega _+\right) \left(\omega _--\omega _+\right)^2}.\label{E:FermiZT}
}
In the right panel of \figref{F:FermiSeeBeck} we plot the linear response figure-of-merit for different on-site energies. 
We find that the figure-of-merit increases exponentially for high temperatures independent of the transition frequencies. In contrast, for a constant chemical potential the figure-of-merit approaches a constant value for high temperatures.
Additionally, we see that there are specific temperatures where the figure-of-merit vanishes. These are the temperatures for which the linear response particle current vanishes and, hence, no power can be extracted.

If the temperature approaches zero, we observe that the figure-of-merit increases again except for the situation where the transition frequencies lie below the Fermi energy of the reservoir. In this case, the figure-of-merit assumes a finite value as the temperature approaches zero. This is again caused by the fact that the relevant energy levels in the reservoirs are occupied for low temperatures.
However, although the conversion of energy into particle currents seems to be very efficient for low and high temperatures, this picture might be misleading since in these regimes the actually generated particle current is exponentially suppressed and so is the power. Thus, one has to look out for a high efficiency at maximum power \cite{Broeck2005,Esposito2009a,Esposito2009b,Broeck2012}. 

It turns out that for the specific situations presented in \figref{F:FermiSeeBeck} the currents become maximal in the interval $T/T_F\in \left[0.13,\,0.23\right]$ (not shown) and, thus, we find that at maximum power the figure-of-merit for the configuration with the transition energies above the Fermi energy is largest with $ZT \sim 12$. Least efficient is the configuration with the lowest transition frequency equal to the Fermi energy. Here, only a figure-of-merit at maximum power of $ZT \sim 0.04$ is reached. Whereas for the configuration with both energies below the Fermi energy we find $ZT \sim 4$ at maximum power. 

Usually, in experiments figure-of-merits of about $ZT \sim 2$ and higher are considered as  efficient. Of course, in our model we assume ideal quantum gases and, thus, the calculated figure-of-merit is probably overestimated. However, from our results we argue that by optimizing the parameters efficiencies at maximum power close or even equal to the optimum are possible. The optimum efficiency in linear response theory is given by half the linear response Carnot efficiency $\eta_{max}=\rklamm{T_{hot}-T_{cold}}/T$ \cite{Esposito2010b}. 
%
%
\section{Ideal Bose Gases}\label{S:BoseGas}
%
%
\begin{figure}[ht]
 \includegraphics[width=0.7\columnwidth]{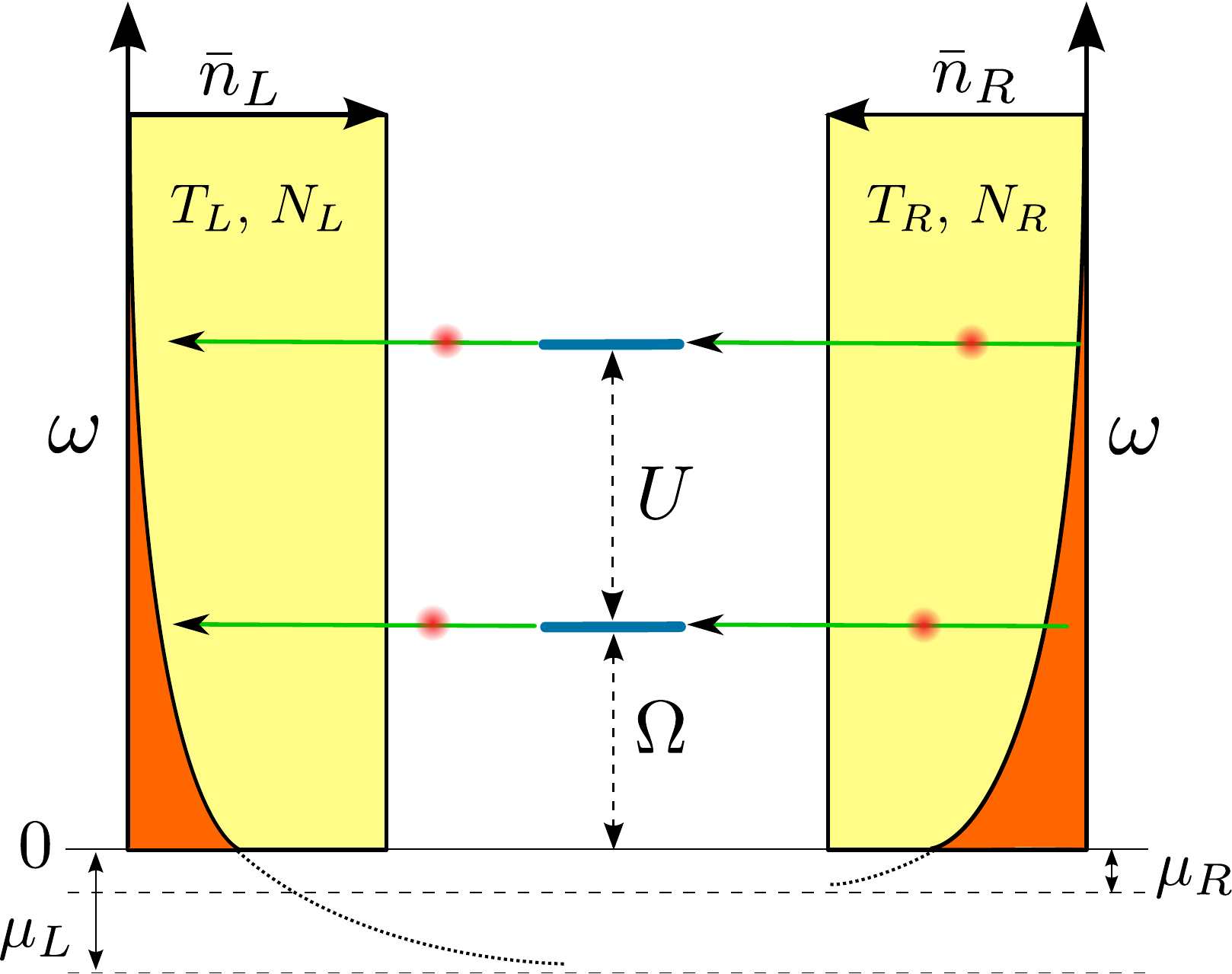}
 \caption{(Color online) Setup for bosonic particle transport. The atomic reservoirs are in thermal equilibrium characterized by temperature $T_\alpha$ and chemical potential $\mu_\alpha=\mu(T_\alpha,N_\alpha)$ for fixed particle number $N_\alpha$. The mean occupation $\bar n_\alpha(\omega)$ of the energy level $\omega$ is given by the Bose-Einstein distribution. The system is described by a harmonic oscillator of frequency $\Omega$ with an additional interaction energy $U$ if two particles are present.}
 \label{F:BoseModel}
\end{figure}

Motivated by the results presented in \secref{S:AtomicReservoirs} we now focus on a transport setup involving bosonic reservoirs. Contrary to the fermionic setup, here, we expect that the critical behavior of the reservoirs leads to characteristic signatures in the transport properties. Therefore, we analyze the transport characteristics of a bosonic transport system as shown in \figref{F:BoseModel}. The bosonic transport system is described by the Hamiltonian
\begin{align}\label{E:SysHamiltonianBose}
 \hat{\mathcal{H}}_{\rm{S}}^{\rm{bose}} &= \frac{U}{2} \hat{a}^\dagger \hat{a} \rklamm{\hat{a}^\dagger \hat{a}-1}+\Omega \hat{a}^\dagger \hat{a}. 
\end{align}
Each particle that is added to or removed from the system changes the system energy at least by a constant amount $\Omega$. If more than one particle is present in the system, these particles interact with a two-body interaction strength $U$ and, thus, increase the total energy of the system. This system Hamiltonian is diagonal in the Fock basis, \ie, $ \hat{\mathcal{H}}_{\rm{S}}^{\rm{bose}}\ket{n}=\omega_n\ket{n}$, with energy eigenvalues $\omega_n=U/2 n(n-1)+\Omega n$. Due to the interaction this energy spectrum is nonlinear and the system generates many non-equivalent transport channels.

The system-bath interaction Hamiltonian reads as
\begin{equation}\label{E:BoseSysBathHamiltonian}
	\hat{H}_{\rm{SB}}=\underset{\alpha,k}{\sum} \left( t_{\alpha,k} \, \hat{b}_{\alpha,k}^{\dagger}\,\hat{a} + \rm{H.\,c.} \right),
\end{equation}
where the tunneling amplitude of an atom hopping from the reservoir $(\alpha)$ into the system or vice versa is proportional to $t_{\alpha,k}^{*}$ and $t_{\alpha,k}$, respectively.
\subsection{Steady-State Current}
In the thermodynamic limit, the steady-state currents through this system measured at reservoir $\alpha$ are given by a sum over all possible system occupations
\equ{
    J_{N}^{(\alpha)} =& - \sum_{n=0}^{\infty} (n+1) \mathcal{I}_n^{(\alpha)},\\
    J_{E}^{(\alpha)} =& \sum_{n=0}^{\infty} (n+1) (\omega_{n}-\omega_{n+1}) \mathcal{I}_n^{(\alpha)},
}
where we defined the abbreviation 
\equ{ 
  \mathcal{I}_n^{(\alpha)}=& \Gamma_\alpha(\omega_{n+1}-\omega_{n})\eklamm{\bar n_\alpha(\omega_{n+1}-\omega_{n})+1} \bar\rho_{n+1}\notag\\
  &-\Gamma_\alpha(\omega_n-\omega_{n+1})\bar n_\alpha(\omega_n-\omega_{n+1}) \bar\rho_n .
  }

Because of the infinite summation, the above expression can not be solved in general. Therefore, we truncate the bosonic Hilbert space at low particle numbers (see Appendix \ref{A:BosonicLiouvillian}). Taking the limit $\lim_{U\rightarrow \infty} \hat{\mathcal{H}}_{\rm{S}}^{\rm{bose}}$  restricts the Hilbert space to at most one bosonic particle in the system . Thus, the system can be either empty or singly occupied which gives rise to a single transition frequency $\Omega$. In this case the steady-state currents in the wide-band limit $\Gamma_{\alpha}(\omega)=\Gamma_{\alpha}$ can be evaluated to
\equ{
      \lim_{U\rightarrow\infty} {J}_{N} = \frac{\Gamma_{\rm{L}} \Gamma_{\rm{R}} \eklamm{\bar n_{\rm{L}}(\Omega)-\bar n_{\rm{R}}(\Omega)}} {\Gamma_{\rm{L}} \eklamm{1+2 \bar n_{\rm{L}}(\Omega)}+\Gamma_{\rm{R}} \eklamm{1+2 \bar n_{\rm{R}}(\Omega)}}, \label{E:SingleWSScurrentBose}
}
which coincides with the result found in Ref.~\cite{Segal2005}.
%
\begin{figure}[t]
 \centering
 \includegraphics[width=\columnwidth]{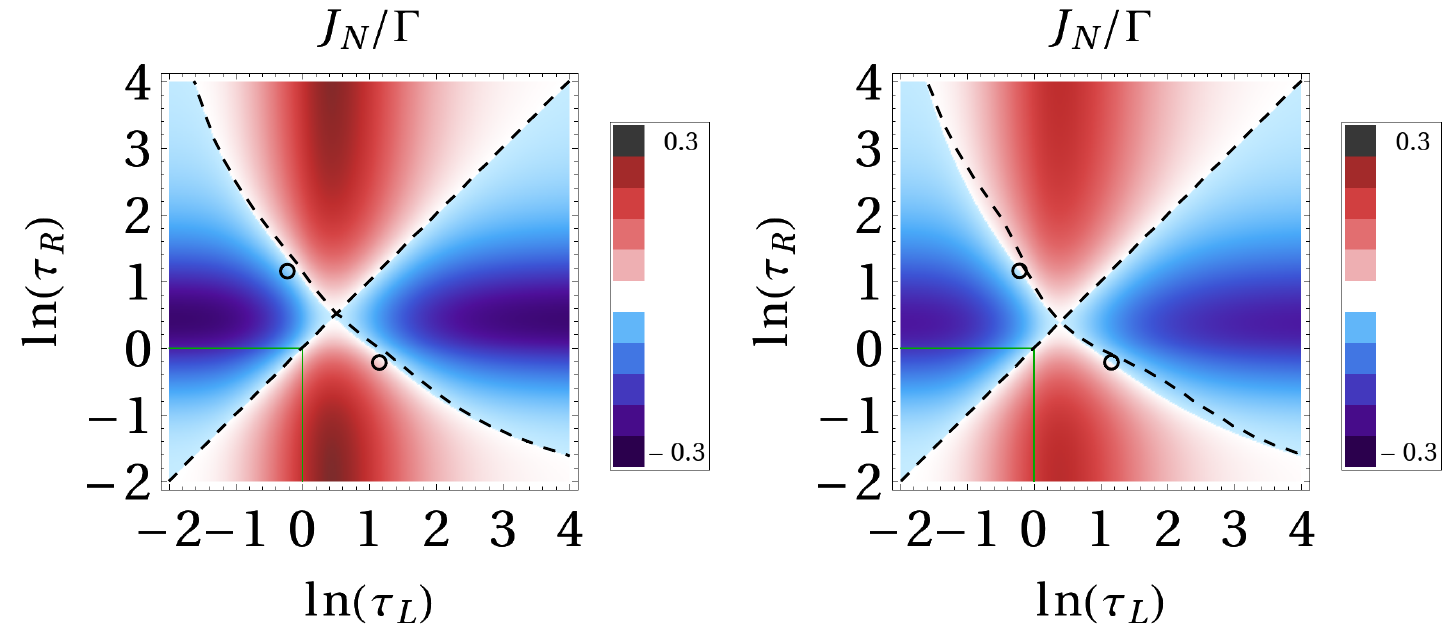}
 \caption{(Color online) Steady-state particle current of the bosonic system versus the dimensionless temperatures of the reservoirs. The  interaction strength is set to $U=0.01 \Omega$ (\textit{left}) and $U=10 \Omega$ (\textit{right}). In both plots, we set $\Omega=T_c$ and $\Gamma_L = \Gamma_R = \Gamma$. The dashed lines indicate a vanishing of the corresponding energy current $J_E$. The solid lines signal the phase transition to a Bose-Einstein condensate in the left and right reservoirs. The circle insets correspond to \figref{F:MeanOccupation}} \label{F:BoseCurrent}
\end{figure}
%
Analogous to the fermionic case we find that the current through a transport channel with energy $\Omega$ is proportional to the difference in the corresponding mean occupations in the left and right reservoir.

As an example we plotted the steady-state particle current for different transition energies in \figref{F:BoseCurrent}. First, we observe that the current is strongest if the interaction strength is weak. Increasing the interaction strength shifts the corresponding transport channel to higher energies which are less occupied in the reservoirs. Therefore, the contribution of these transport channels to the current is diminished.
Additionally, we see two lines where the steady-state particle current vanishes. The diagonal line reflects the thermodynamic equilibrium, \ie if $\Delta_T=\Delta_n=0$. The reason for the emergence of the second line lies in the temperature dependence of the mean occupations as discussed for the fermionic setup in \secref{S:FermiSteadyStateCurrent}.

For the energy current (not shown), we find qualitatively the same behavior as for the particle current. However, depending on the system parameters the energy current can be finite even for vanishing particle current. We indicated the temperatures where the energy current vanishes by dashed lines in both plots of \figref{F:BoseCurrent}. We observe that the nonequilibrium lines where the energy current vanishes are shifted to higher temperatures compared to the vanishing particle current. 

Contrary to the fermionic steady-state current plotted in \figref{F:FermiCurrent}, we do not observe a qualitative change in the bosonic particle current in dependence of the transition energies. This behavior stems from the fact that there is no equivalent of the Fermi energy and no Pauli exclusion principle in bosonic systems. Hence, the bosonic mean occupations look qualitatively the same for all energy levels (see \figref{F:MeanOccupation}).

Within the condensate phase, we observe a finite particle-current which results from the thermal fraction of the Bose gas. This thermal fraction decreases with temperature like $T^{3/2}$ and, therefore, the current exactly vanishes at $T=0$.

In our BMS master-equation approach, the coherences decouple from the occupations and, thus, can be neglected. However, if one enters the condensate phase the coherences between the particles become stronger with decreasing temperature. Therefore, the decoupling between coherences and occupations is not expected to hold and the coherences can not be neglected anymore. Hence, we do not expect that our results remain valid deep in the condensate phase. 
\subsection{Transport Coefficients}
In the limit of a single transition-frequency in the system, \ie, $U=0$ or $U\rightarrow\infty$, the energy current is proportional to the particle current. As shown in \secref{S:TransportCoefficients} in this situation not all transport coefficients can be calculated. Therefore, we consider the case of two transport channels with different energies. This situation is established by truncating the Hilbert space at two particles leading to two transitions in the system from zero to one particle, \ie $\omega_{1}=\Omega$ and from one to two particles, \ie, $\omega_{2}=U+\Omega$. In the following paragraphs, we present the resulting bosonic transport coefficients in the wide-band limit.
Some results for the bosonic transport coefficients are plotted in Figs.~\ref{F:BoseConductance} and \ref{F:BoseSeebeck}. 

\subsubsection{Matter Conductance}
The bosonic matter conductance for the considered transport setup reads as
\equ{
    \tilde\sigma = \frac{\bar\Gamma \,\bar{n}(\omega _1) \eklamm{1+\bar{n}(\omega _2)} \eklamm{1+\bar{n}(\omega _1)+2 \bar{n}(\omega _2)}}{T \sklamm{1+\bar{n}\left(\omega _2\right)+\bar{n}\left(\omega _1\right) \eklamm{2+3\bar{n}\left(\omega _2\right)}} }.\label{E:BoseDiffusion}
}
%

\begin{figure}
 \centering
 \includegraphics[width=\columnwidth]{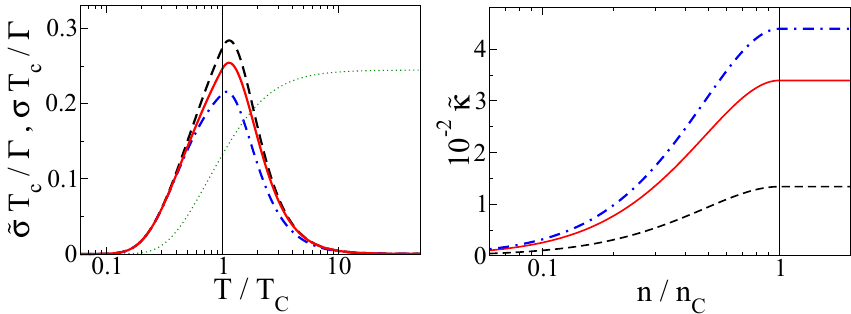}
 \caption{(Color online) \textit{Left}: Matter conductance for the ideal Bose gas reservoirs plotted for different on-site interaction strength $U=0.5 \Omega$ (dashed line), $U= \Omega$ (solid line) and $U=\infty$ (dotted-dashed line) versus the inverse normalized temperature. The transition energy is fixed to $\Omega= T_c$. For a constant chemical potential $\mu=-0.5 T_c$ (thin, dotted line) where we use $U=\Omega=T_c$, the matter conductance becomes constant for high temperatures. \textit{Right}: Thermal conductance for the ideal Bose gas reservoirs plotted for different on-site interaction strength $U=0.5 \Omega$ (dashed line), $U= \Omega$ (solid line) and $U=3 \Omega$ (dotted-dashed line) versus the inverse normalized density. The single-particle energy is $\Omega= T$ and the rates are fixed to $\Gamma_L = \Gamma_R = \Gamma$.  }\label{F:BoseConductance}
\end{figure}
%
In the left panel of \figref{F:BoseConductance}, we plot this transport coefficient for different transition energies versus the normalized temperature. 
Depending mostly on the lowest transition frequency $\Omega$, the matter conductance has a maximum value at a finite temperature above the critical value $T_c$. Decreasing the lower transition frequency shifts the maximum closer to the critical temperature whereas increasing the transition frequency shifts the maximum away from the critical temperature. The maximum can never lie below the critical temperature since there the chemical potential vanishes and the number of thermal particles which can contribute to the particle current decreases. 

Increasing the temperature leads to a decrease of the matter conductance since the occupation of the transition energy level in the reservoirs is reduced. The influence of the second transport channel is mainly reflected in the maximum value of the transport coefficient. This value is increased if the transport channels are close together, \ie, if the interaction strength $U$ is small. If the interaction strength is increased the respective transport channel is shifted to higher energies and contributes less to the current because of the lower occupations in the reservoir. Thus, the maximum conductance decreases to a minimum value resulting from the lower transport channel  (dotted-dashed line). 

For comparison we also included a plot for the conductance with constant chemical potential $\mu=-0.5 T_c$ (dotted line). Here, the conductance takes on a constant finite value $ $ in the limit of high temperature. This is caused by the fact that for a constant chemical potential the occupations of energy levels in the reservoirs increase linearly with the temperature in the high-temperature limit. If the temperature approaches zero, the conductance vanishes due to the depletion of the transition energy-levels in the reservoirs.
\subsubsection{Heat Conductance}
The bosonic heat conductance for the considered transport setup reads as
\equ{
    \tilde\kappa = \frac{2 \left[1+\bar{n}\left(\omega _1\right)\right] \bar{n}\left(\omega _2\right) \left(\omega _1-\omega _2\right)^2}{T \left[1+\bar{n}\left(\omega _1\right)+2 \bar{n}\left(\omega _2\right)\right]^2} \tilde\sigma.\label{E:BoseThermalConductance}
}
In the right panel of \figref{F:BoseConductance}, we plot the thermal conductance for different transition energies versus the  normalized density. We observe that this transport coefficient increases with increasing density and reaches a maximum value at the critical density when Bose-Einstein condensation sets in. The value of the maximum depends on the transition energies of the system. In general, there is a finite interaction strength that maximizes the heat conductance. For a low interaction-strength, the heat conductance is strongly decreased since it is proportional to $U^2$. For a high interaction-strength, the heat conductance is also diminished because
the occupation of the upper transition energy is decreased.

For densities above the critical value, the thermal conductance remains constant since the thermal gas fraction in this phase is independent of the density and depends on the temperature only. All additional particles occupy the reservoir ground state and, thus, do not contribute to the currents.
\subsubsection{Thermopower}
The analog to the thermopower for the considered bosonic transport setup in the wide-band limit looks formally the same as for the fermionic case
\equ{
    \tilde\Sigma = \frac{\braket{\phi_b(\omega_1,\omega_2)}}{T}.\label{E:BoseSeebeck}
}
However, here appears the bosonic average energy which we define as
\equ{
    \braket{\phi_b(\omega_1,\omega_2)}=\frac{\left[1+\bar{n}\left(\omega _1\right)\right] \phi\left(\omega _1\right)+2 \bar{n}\left(\omega _2\right) \phi\left(\omega _2\right)}{1+\bar{n}\left(\omega _1\right)+2 \bar{n}\left(\omega _2\right)},\label{E:BoseMeanTransportedEnergy}
}
where we used the expression defined in \equref{E:SingleParticleHeat}. In the left panel of \figref{F:BoseSeebeck}, we plot the temperature dependence of this transport coefficient for different values of the interaction strength. Analogously to the fermionic case, we find that the modified Seebeck coefficient takes on a finite positive value in the high temperature limit where the average transported energy become $3/2 k_B T$.

When the temperature is lowered, the modified thermopower decreases. At a temperature where the chemical potential contribution starts to dominate over the transport-channel energies, the modified thermopower changes its sign. 
When the temperature is decreased further, the modified thermopower crosses the critical temperature of the phase transition continuously. However, at the critical temperature the modified thermopower is not analytic. Thus, the second derivative with respect to temperature shows a jump when the condensate phase is entered. This behavior is also well known for the heat capacity of the ideal Bose gas \cite{Pitaevskii}. 

In the condensate phase, the modified thermopower decreases further and diverges like $-1/T$ when the temperature is close to absolute zero. In general, the particle current is mainly influenced by the lower transport channel. Hence, the modified thermopower just weakly depends on the interaction strength $U$. For high and low values of the interaction strength, the modified thermopower approaches the single-frequency limit result (dashed line). In-between, there is a finite interaction-strength that maximizes the modified thermopower at the critical temperature (solid line). However, the relative increase in the thermopower output is still small.

On the contrary, the approach with a constant chemical potential predicts a vanishing thermopower for high temperatures (dotted line). There is no change of sign of the thermopower in dependence of the reservoir temperature. Additionally, the conventional thermopower is continuous and differentiable at the critical temperature and, thus, it is not sensitive to the quantum phase transition of the ideal Bose gas. 
\subsubsection{Figure-of-merit}
Finally, we analyze the efficiency of the bosonic thermodynamic device characterized by the figure-of-merit which for the bosonic system reads as
%
\begin{figure}
 \centering
 \includegraphics[width=\columnwidth]{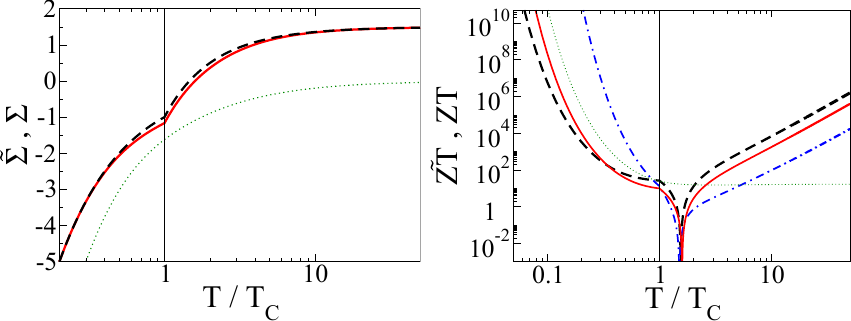}
 \caption{(Color online) \textit{Left}: Plot of the thermopower for different interaction strengths $U= \Omega$ (solid line) and $U=\infty$ (dashed line). \textit{Right}: Plot of the figure-of-merit versus the normalized temperature for different interaction strengths $U=0.5 T_c$ (dashed line), $U= T_c$ (solid line) and $U=5 T_c$ (dotted-dashed line). For $U=T_c$ and a constant chemical potential $\mu=-0.5 T_c$ (thin, dotted line), the thermopower vanishes for high temperatures and the figure-of-merit becomes constant. The transition energy is fixed to $\Omega= T_c$ in all plots.}\label{F:BoseSeebeck}
\end{figure}
%
\equ{
    \tilde{ZT} =\frac{\left[1+\bar{n}\left(\omega _1\right)+2 \bar{n}\left(\omega _2\right)\right]^2 \braket{\phi_b(\omega_1,\omega_2)}^2}{2 \left[1+\bar{n}\left(\omega _1\right)\right] \bar{n}\left(\omega _2\right) \left(\omega _1-\omega _2\right){}^2}.\label{E:BoseZT}
}
In the right panel of \figref{F:BoseSeebeck}, we plot the temperature dependence of the figure-of-merit for different values of the interaction strength. We observe that the figure-of-merit vanishes for vanishing linear response particle current. At the critical temperature of the phase transition, the figure-of merit is nondifferentiable but continuous. In the limit of high and low temperatures, the figure-of-merit increases exponentially. 

However, if we compare it with the linear response particle current (not shown), we find that the figure-of-merit at maximum negative current around $T/T_c\sim2.3$ takes on the values $\tilde{ZT}\sim 34$ (dashed line), $\tilde{ZT}\sim 8$ (solid line) and $\tilde{ZT}\sim 1$ (dotted-dashed line) in the thermal phase. Additionally, the linear response current shows a positive maximum around $T/T_c\sim0.6$  in the condensate phase, where we find that the figure-of-merit takes on the values $\tilde{ZT}\sim 34$ (dashed line),  $\tilde{ZT}\sim 15$ (solid line) and $\tilde{ZT}\sim 152$ (dotted-dashed line). 

For comparison, we also plot the figure-of-merit for a constant chemical potential (thin, dotted line). Again, we find that in this case the high-temperature behavior is modified as the figure-of-merit becomes constant.
%
%
\section{Summary}\label{S:Summary}
%
%
We calculated the steady-state fluxes and affinities from a master equation in Born-Markov-Secular approximation for a general two-terminal transport setup. There, we took into account that in transport experiments with ultracold atoms the chemical potential in general depends on the temperature and the particle density of the reservoirs.

We found that the nonlinearity introduced by the temperature- and density-dependent chemical potential strongly influences the steady-state particle and energy currents through the system. Depending on the energy spectrum of the system, we could observe multiple regimes where the steady-state currents flow with or against an externally applied thermal bias. This counterintuitive result stems from the temperature- and density-dependencies of the mean occupations in the reservoirs induced by the chemical potential.

Subsequently, we derived the corresponding Onsager system of equations from which we calculated the linear response transport coefficients. Comparing the results for fermionic and bosonic transport, we found clear signatures of criticality in the bosonic transport coefficients. Thus, transport measurements provide new tools to study critical phenomena in nonequilibrium setups.

Finally, we investigated the figure-of-merit for the bosonic and fermionic setups. In correspondence with experimental results \cite{Brantut2013}, we found that high figures-of-merit at maximum power can be obtained in both systems. This suggests to further investigate transport setups using ultracold atomic gases in view of efficient thermopower devices.

Financial support by the DFG (SFB 910, SCHA 1646/2-1 and GRK 1558) is gratefully acknowledged.

\appendix
%
%
\section{Onsager Matrices Transformation}\label{A:OnsagerMatrices}
%
%
Depending on the experimental setup, different intensive state parameters can be held constant. Thus, the experimentally controllable affinities change accordingly. For the setups compared within this paper, we focus on situations with constant chemical potential where the affinities are given by $\Delta_\mu$ and $\Delta_T$. On the other hand, we analyze a setup with constant particle density where the affinities are given by $\Delta_n$ and $\Delta_T$. These constraints yield different affinities which in linear response theory can be related to each other according to the linear transformation
\begin{align}
  \left(
    \begin{array}{c}
       \Delta _{\beta } \\
       \beta  \Delta _{\mu }
    \end{array}
  \right)
  =
  \mathbb{A}^{\rm{T}}\cdot
  \left(
    \begin{array}{c}
       \Delta _{\beta } \\
        \beta \frac{\partial\mu}{\partial n}\Delta_n
    \end{array}
  \right),\; \mathbb{A}^{\rm{T}}=\left(
    \begin{array}{cc}
       1 & 0 \\
       \beta \frac{\partial\mu}{\partial\beta} & 1
    \end{array}
  \right).\label{E:AffinityTransformation}
\end{align}
Here, we used the linear expansion of the potential bias $\Delta_\mu$ with respect to the new affinities $\Delta_n$ and $\Delta_\beta$ which reads as
\equ{\Delta_\mu = \frac{\partial\mu}{\partial n} \Delta_n + \frac{\partial\mu}{\partial \beta} \Delta_\beta.}
Analogously, we find that the generalized linear fluxes can be transformed according to 
\begin{align}
   \left(
    \begin{array}{c}
       -\tilde{J}_Q \\
       {J}_N
    \end{array}
  \right)
  =
  \mathbb{A}\cdot
  \left(
    \begin{array}{c}
       -J_Q \\
       J_N
    \end{array}
  \right).
\end{align}
Due to the linearity of the system of equations we can also find a transformation for the Onsager matrices themselves. Inserting the Onsager system \eqref{E:UsualOnsagerSystem} together with \eqref{E:AffinityTransformation} we find 
\begin{align}
  {\boldsymbol{\tilde M}}
  =
  \mathbb{A}\cdot{\boldsymbol{M}}\cdot\mathbb{A}^{\rm{T}},
\end{align}
which yields the relation stated in \equref{E:OnsagerCorrespondence}. With this, results obtained for a setup with constant chemical potential can be transformed to the corresponding result for the case of constant particle density.
%
%
\section{Fermionic Liouvillian}\label{A:FermionicLiouvillian}
%
%
The reduced density matrix of an electronic double quantum dot in the Coulomb blockade regime has the diagonal matrix elements $\rho_0=\ket{0}\bra{0}$, $\rho_-=\ket{-}\bra{-}$ and $\rho_+=\ket{+}\bra{+}$. Using the wide-band limit $\Gamma_\alpha(\omega)=\Gamma_\alpha$ the conditioned Liouvillian \eqref{E:LiouvillianChiEta} in the energy eigenbasis which obeys 
\begin{equation}
  \frac{d}{dt}
  \begin{pmatrix}
   \rho_0 \\ \rho_- \\ \rho_+
  \end{pmatrix}
  =
  \mathcal{L}(\boldsymbol{\chi},\boldsymbol{\eta})
  \begin{pmatrix}
   \rho_0 \\ \rho_- \\ \rho_+
  \end{pmatrix},
\end{equation}
is given by
\begin{widetext}
  \begin{align}
    \mathcal{L}(\boldsymbol{\chi},\boldsymbol{\eta})=\frac{1}{2}\underset{\alpha \in \sklamm{L,R}}{\sum}\Gamma_\alpha \left(
      \begin{array}{ccc}
	  \bar{n}_{\alpha }(\omega_-) +\bar{n}_{\alpha }(\omega_+)  & - e^{\ii \left(\chi _{\alpha }-\eta _{\alpha } \omega _-\right)} \left[ 1-\bar{n}_{\alpha }\left(\omega _-\right)\right] & -   e^{\ii \left(\chi _{\alpha }-\eta _{\alpha } \omega _+\right)} \left[ 1-\bar{n}_{\alpha }\left(\omega _+\right)\right] \\
	 - e^{-\ii \left(\chi _{\alpha }-\eta _{\alpha } \omega _-\right)}  \bar{n}_{\alpha }\left(\omega _-\right) &  1-\bar{n}_{\alpha }\left(\omega _-\right) & 0 \\
	 -   e^{-\ii \left(\chi _{\alpha }-\eta _{\alpha } \omega _+\right)} \bar{n}_{\alpha }\left(\omega _+\right) & 0 &  1-\bar{n}_{\alpha }\left(\omega _+\right)
      \end{array}
      \right).
  \end{align}
\end{widetext}
The steady-state vector $\bar\rho=\rklamm{\bar\rho_0,\bar\rho_-,\bar\rho_+}^{\rm{T}}$ of this Liouvillian is defined by $\mathcal{L}(\boldsymbol{0},\boldsymbol{0})\bar\rho=0$ and reads as
\begin{equation}
     \bar\rho
     =
     \frac{1}{\theta}
     \underset{\alpha,\beta \in\sklamm{L,R}}{\sum} \Gamma_\alpha \Gamma_\beta
     \begin{pmatrix}
      \eklamm{1-\bar n_\alpha (\omega_-)} \eklamm{1-\bar n_\beta (\omega_+)} \\
      \bar n_\alpha (\omega_-) \eklamm{1-\bar n_\beta (\omega_+)} \\
      \eklamm{1-\bar n_\alpha (\omega_-)} \bar n_\beta (\omega_+)
     \end{pmatrix},
\end{equation}
where the normalization factor $\theta$ is given by
\equ{\theta = \rklamm{\Gamma_L+\Gamma_R}^2-\underset{\alpha,\beta \in\sklamm{L,R}}{\sum} \Gamma_\alpha \Gamma_\beta \bar n_\alpha (\omega_-) \bar n_\beta (\omega_+) .}
With this result, we can calculate the fermionic energy and particle currents according to Eqs.\ (\ref{E:StaticCurrentDefinition}) and (\ref{E:StaticEnergyCurrentDefinition}).

%
%
\section{Bosonic Liouvillian}\label{A:BosonicLiouvillian}
%
%

In order to solve the master equation for the bosonic system, we have to truncate the respective Hilbert space at low particle numbers. For a bosonic system with two transition frequencies only, we truncate the bosonic Hilbert space described by the Hamiltonian \eqref{E:SysHamiltonianBose} such that at most two particles at a time can be present in the system. Then the eigenstates are given by the bosonic Fock states $\ket{0}$, $\ket{1}$ and $\ket{2}$ with the corresponding eigenvalues $\omega_0=0$, $\omega_1=\Omega$ and $\omega_2=\Omega+U$.

Using the wide-band limit $\Gamma_\alpha(\omega)=\Gamma_\alpha$ the conditioned bosonic Liouvillian \eqref{E:LiouvillianChiEta} in the energy eigenbasis which obeys 
\begin{equation}
  \frac{d}{dt}
  \begin{pmatrix}
   \rho_0 \\ \rho_1 \\ \rho_2
  \end{pmatrix}
  =
  \mathcal{L}(\boldsymbol{\chi},\boldsymbol{\eta})
  \begin{pmatrix}
   \rho_0 \\ \rho_1 \\ \rho_2
  \end{pmatrix},
\end{equation}
is given by
\begin{widetext}
  \begin{align}
    \mathcal{L}(\boldsymbol{\chi},\boldsymbol{\eta})= \underset{\alpha \in \sklamm{L,R}}{\sum} \Gamma_\alpha \left(
      \begin{array}{ccc}
	  -\bar{n}_{\alpha }(\omega_1) & e^{\ii \left(\chi _{\alpha }-\eta _{\alpha } \omega _1\right)} \left[ 1+\bar{n}_{\alpha } (\omega _1)\right] & 0 \\
	 e^{-\ii \left(\chi _{\alpha }-\eta _{\alpha } \omega _1\right)}  \bar{n}_{\alpha }\left(\omega _1\right) &  -2 \bar n_\alpha (\omega_2) - \left[ 1+\bar{n}_{\alpha } (\omega _1)\right] & 2 e^{\ii \left(\chi _{\alpha }-\eta _{\alpha } \omega _2\right)}\eklamm{1+\bar{n}_{\alpha }\left( \omega _2 \right)} \\
	 0 & 2 e^{-\ii \left(\chi _{\alpha }-\eta _{\alpha } \omega _2\right)} \bar n_\alpha (\omega_2) & -2\eklamm{1+\bar{n}_{\alpha }\left( \omega _2 \right)}
      \end{array}
      \right).
  \end{align}
\end{widetext}
The steady-state vector $\bar\rho=\rklamm{\bar\rho_0,\bar\rho_1,\bar\rho_2}^{\rm{T}}$ of this Liouvillian is defined by $\mathcal{L}(\boldsymbol{0},\boldsymbol{0})\bar\rho=0$ and reads as
\begin{equation}
     \bar\rho
     =
     \frac{1}{\theta}
     \underset{\alpha,\beta \in\sklamm{L,R}}{\sum} \Gamma_\alpha \Gamma_\beta
     \begin{pmatrix}
      \eklamm{1+\bar n_\alpha (\omega_1)} \eklamm{1+\bar n_\beta (\omega_2)} \\
      \bar n_\alpha (\omega_1) \eklamm{1+\bar n_\beta (\omega_2)} \\
      \bar n_\alpha (\omega_1) \bar n_\beta (\omega_2)
     \end{pmatrix},
\end{equation}
where the normalization factor $\theta$ is given by
\equ{\theta = \underset{\alpha,\beta }{\sum} \Gamma_\alpha \Gamma_\beta \sklamm{1+\bar n_\beta (\omega_2) + \bar n_\alpha (\omega_1)\eklamm{2+3 \bar n_\beta(\omega_2)}} .}
With this result we can calculate the bosonic energy and particle currents according to Eqs.\ (\ref{E:StaticCurrentDefinition}) and (\ref{E:StaticEnergyCurrentDefinition}).
%
%
%
%
\bibliographystyle{apsrev}

\begin{thebibliography}{55}
\bibitem[{Aff()}]{Affinity}
\bibinfo{note}{The derivative $ \partial\mu/\partial n$ corresponds to the
  inverse compressibility of the ideal Bose gas in equilibrium. Thus, one could
  also choose a generalized energy current $\tilde J_n = \partial\mu/\partial n
  J_N$ with the corresponding affinity $\beta \Delta_n$. This would introduce
  an additional prefactor $\partial\mu/\partial n$ in the kinetic coefficient
  $\tilde L_{12}$ and a prefactor $(\partial\mu/\partial n)^2$ in $\tilde
  L_{22}$. The transport coefficients $\tilde\sigma$ and $\tilde \Sigma$ now
  depend on the compressibility and change their physical meaning.}
\end{thebibliography}

\end{document}